\documentclass[nofootinbib,prb,reprint,twocolumn,amsmath,amssymb]{revtex4-1}

\newcommand{\shp}[2]{N^{#1}_{#2}(\bx)}

\newcommand{\del}{\nabla}

\newcommand{\bc}{\boldsymbol{\textbf{c}}}
\newcommand{\bd}{\boldsymbol{\textbf{d}}}

\newcommand{\bx}{\boldsymbol{\textbf{x}}}
\newcommand{\bA}{\textbf{A}}

\newcommand{\bH}{\textbf{H}}

\newcommand{\bQ}{\textbf{Q}}

\newcommand{\bM}{\textbf{M}}

\newcommand{\bR}{\boldsymbol{\textbf{R}}}
\newcommand{\bRIx}{\boldsymbol{\textbf{R}}^{\bx}_I}

\newcommand{\bY}{\textbf{Y}}

\newcommand{\intomega}{\int_{\Omega}}
\newcommand{\intomegai}{\int_{\Omega_{I}}}

\newcommand{\dx}{\,d\bx}

\newcommand{\Rthree}{ {\mathbb{R}^{3}} }

\newcommand{\bk}{\boldsymbol{\textbf{k}}}
\newcommand{\vecc}{\mathbf{c}}

\newcommand{\phitot}{\phi_{\text{tot}}}

\newcommand{\bsmear}{b^{\text{smear}}}

\newcommand{\phiAux}{\phi_{\text{aux}}}
\newcommand{\ualbk}{u_{\alpha,\mathbf{k}}}
\newcommand{\psialbk}{\psi_{\alpha,\mathbf{k}}}
\newcommand{\epsalbk}{\epsilon_{\alpha,\mathbf{k}}}
\usepackage{subfiles}
\usepackage{graphicx}
\usepackage{graphics}
\usepackage{multirow}
\usepackage{fancyhdr, ifpdf}
\usepackage{amsxtra}
\usepackage{stmaryrd}
\usepackage{mathrsfs}
\usepackage{mathtools}
\usepackage{psfrag}
\usepackage{threeparttable}
\usepackage{subfigure}
\usepackage{epsfig}
\usepackage{rotating}
\usepackage{setspace}
\usepackage{float}
\usepackage{amsfonts}
\usepackage{amsbsy}
\usepackage{amscd}
\usepackage{amsthm}
\usepackage{color}
\usepackage{tabularx}
\usepackage{sidecap}
\usepackage{dcolumn}
\usepackage{footnote}
\usepackage{algorithm}
\usepackage{algorithmic}
\usepackage{enumitem}
\usepackage[titletoc,title]{appendix}

\usepackage[normalem]{ulem}
\usepackage{soul}

\definecolor{hellgruen}{rgb}{0.2,0.7,0.2}

\newcolumntype{M}[1]{>{\centering\arraybackslash}m{#1}}
\newcolumntype{N}{@{}m{0pt}@{}}

\graphicspath{ {./images/} }

\begin{document}
\raggedbottom
\title{Fast and robust all-electron density functional theory calculations in solids using orthogonalized enriched finite elements}
\author{Nelson D. Rufus}
\affiliation{Department of Mechanical Engineering, University of Michigan, Ann Arbor, Michigan 48109, USA}
\author{Bikash Kanungo}
\affiliation{Department of Mechanical Engineering, University of Michigan, Ann Arbor, Michigan 48109, USA}
\author{Vikram Gavini}
\affiliation{Department of Mechanical Engineering, University of Michigan, Ann Arbor, Michigan 48109, USA}
\affiliation{Department of Materials Science and Engineering, University of Michigan, Ann Arbor, Michigan 48109, USA}
\begin{abstract}
We present a computationally efficient approach to perform systematically convergent real-space all-electron Kohn-Sham DFT calculations for solids using an enriched finite element (FE) basis. The enriched FE basis is constructed by augmenting the classical FE basis with atom-centered numerical basis functions, comprising of atomic solutions to the Kohn-Sham problem. Notably, to improve the conditioning, we orthogonalize the enrichment functions with respect to the classical FE basis, without sacrificing the locality of the resultant basis. In addition to improved conditioning, this orthogonalization procedure also renders the overlap matrix block-diagonal, greatly simplifying its inversion. Subsequently, we use a Chebyshev polynomial based filtering technique to efficiently compute the occupied eigenspace in each self-consistent field iteration. We demonstrate the accuracy and efficiency of the proposed approach on periodic unit-cells and supercells. The benchmark studies show a staggering $130\times$ speedup of the orthogonalized enriched FE basis  over the classical FE basis. We also present a comparison of the orthogonalized enriched FE basis with the LAPW+lo basis, both in terms of accuracy and efficiency. Notably, we demonstrate that the orthogonalized enriched FE basis can handle large system sizes of $\sim$10,000 electrons. %\tcr{Furthermore, for the benchmark systems under consideration, we attain a scaling of $\mathcal{O}(N_e^{2.3})$ ($N_e$ being the number of electrons), as opposed to $\mathcal{O}(N_e^{2.8})$ for LAPW+lo.}
Finally, we observe good parallel scalability of our implementation with $92\%$ efficiency at $22\times$ speedup for a system with 620 electrons.
%for a 62 silicon carbide (SiC) divacany system.
\end{abstract}
\maketitle

%%%%%%%%%%%%%%%%%%%%%%%%%%%%%%  INTRO %%%%%%%%%%%%%%%%%%%%%%%%%%%%%%%%%%%%%%%%
%\input{intro}
\section{\label{sec:intro}Introduction}
Density Functional Theory (DFT) has been the workhorse of electronic structure calculations over the past several decades. The theory states that all ground-state properties of materials can be completely determined from the ground-state electron density~\cite{dftHohKohn}. One of the most common methods to construct this density is by using the Kohn-Sham formulation~\cite{dftKohnSham} which replaces the many-body problem with a single electron problem in an effective potential. The many body interactions are encapsulated in one component of this potential---the exchange-correlation potential. While the exact form of this potential is unknown, several approximations are available~\cite{richardMartin}. Beyond the exchange-correlation approximation, typical DFT calculations also employ a pseudopotential approximation~\cite{pspPayne,pspBachelet,pspChelik,pspHamann,pspSchwerdtfeger}, to attain a good balance of computational efficiency and accuracy. To elaborate, the pseudopotential models the effect of the singular nuclear potential and the core electrons into a smooth effective potential. As a result, it simplifies the Kohn-Sham problem to the evaluation of only the smooth pseudo-wavefunctions corresponding to the valence electrons. Despite the success and widespread use of pseudopotentials, some numerical studies over the past two decades have highlighted the limitations of the pseudopotential approximation. Some of them include the study of ground-state properties of compounds of inner-transition metals~\cite{fabris2005taming,kresse2005comment}, phase transition properties of semiconductors\cite{abu2000fp,xiao2010first} and transition metal oxides~\cite{kolorenvc2007bmno}, ionization potentials of actinide atoms~\cite{liu1998ab}, point defects in refractory metals~\cite{fernandez2020energeticptdefectmetal}, excited state properties with many-body perturbation theory~\cite{gomez2008influence,govoni2018gw100}, etc. %\sout{As a result, the more accurate all-electron approach to DFT remains critical for such cases where the pseudopotentials do not provide the desired accuracy.} 
Although substantial recent progress has been made with the advent of multi-projector pseudopotential formulations\cite{oncv},  all-electron calculations serve as a useful avenue for systems where pseudopotentials lack in accuracy and also aid pseudopotential transferability studies.

Although all-electron calculations provide for a complete description of the materials system, they come at a substantially high computational cost, owing to the numerical challenge in capturing the sharp variations of the electronic fields and the need to compute for much larger number of single electron states. Historically, all-electron calculations have been conducted using atom-centered orbitals\cite{FHI_BLUM20092175, crystalpaper, NWchem-VALIEV20101477}. This entails the use of a few atom-specific basis functions per atom, and thereby afford good computational efficiency. However, owing to the incomplete nature of the basis, they lack systematic convergence and may not provide the desired accuracy, especially for metallic systems~\cite{Jensen1GaussianBasis, Jensen2GaussianBasis, FellerGaussianBasisMouse}. The other widely used approach to all-electron DFT calculations involves the augmented planewave~\cite{slater_apw} family of methods, which includes the augmented planewave (APW) ~\cite{Loucks1967, Koelling1975}, linearized augmented planewave (LAPW) ~\cite{andersen1975linear, Wimmer1981, Weinert1982}, APW+lo (localized orbitals)~\cite{Sjostedt2000, Madsen2001, Gulans2014}, and LAPW+lo~\cite{singh1991lo,schwarz2002electronicwein2k} methods. In these methods, the simulated physical domain is divided into two regions: atom centered spheres called muffin tins (MTs) and the interstitial region. The basis functions in the interstitial region are planewaves. Inside the MTs, the basis functions are products of radial functions and spherical harmonics. The radial functions are solutions to the 1D radial Kohn-Sham equation, solved using a spherically averaged potential and a choice of an energy parameter. 
While the planewave augmented methods are efficient for all-electron calculations, the quality of the basis remains sensitive to choice of the MT radius, the  core-valence split, the function matching constraints at sphere boundary, and the energy parameter used in constructing the radial functions. Moreover, with the usage of planewaves in the interstitial regions they inherit certain notable disadvantages of planewaves, such as their restrictions to periodic boundary conditions and the limited parallel scalability owing to the the extended nature of planewaves.

An alternative approach that has recently gained prominence for DFT calculations is the finite element (FE) method\cite{bathe2006finite}, which comprises of local piecewise continuous polynomials. Like planewaves, the FE basis is complete, and provides systematic convergence. However, unlike planewaves, the FE basis offers additional advantages of locality that affords good parallel scalability, ease of adaptive spatial resolution, and the ability to handle arbitrary boundary conditions. In the context of pseudopotential calculations, there exists a growing body of works~\cite{White1989, Tsuchida1998, Pask1999, Pask2001, Pask2005, Zhang2008, Suryanarayana2010, Fang2012, Bao2012, Motamarri2013, Das2019,Motamarri2020} that establishes usefulness of the FE basis. Particularly, recent efforts~\cite{Das2019,Motamarri2020} at efficient FE based DFT calculations have outperformed planewaves by $5-10\times$, for pseudopotential calculations beyond system sizes containing a few
hundred atoms. For all-electron calculations, although some of the works~\cite{White1989, Tsuchida1996, Batcho2000, Bylaska2009, Lehtovaara2009, Alizadegan2010, Bao2012, Motamarri2013, Schauer2013, Motamarri2014, Maday2014, Davydov2016} have demonstrated the promise of the FE basis, the efficiency of the FE basis remains unsatisfactory when compared to the atomic orbitals type basis. As shown in~\cite{Motamarri2013}, with regards to all-electron calculations, the FE basis is an order of magnitude slower than the gaussian basis. 

An enriched finite element (EFE) basis, wherein the classical FE (CFE) basis (i.e., the standard FE basis) is augmented with atom-centered basis, termed as \emph{enrichment functions}, offers a way to greatly improve the efficiency of the FE basis. Several efforts have explored the efficacy as well as the various numerical aspects of employing an EFE basis for DFT calculations. Previous efforts have explored the EFE basis for the solution of the Schr\"odinger and the Kohn-Sham equations in the context of pseudopotential  calculations~\cite{Sukumar2009, Pask2011, Pask2017} as well as the electrostatic problem arising in all-electron calculations~\cite{ Pask2011, pask2012linear}. In these works, the size of EFE basis required to reach chemical accuracy was shown
to be an order of magnitude smaller than the corresponding planewave
basis and two orders of magnitude smaller than the corresponding
CFE basis~\cite{Pask2011, Pask2017}. In the context of the full ground-state all-electron calculations, the promise of an EFE basis was, first, established by combining the CFE basis with the standard gaussian basis~\cite{Yamakawa2005} (see ~\cite{Pritchard2019} for more about the standard gaussian basis). More recently, in ~\cite{kanungo2017large} a more efficient EFE basis for all-electron calculations have been proposed by combining the CFE basis with numerical atom-centered basis. Given that the enrichment functions are extended in space, maintaining the locality of the resultant basis as well as the sparsity of the discrete matrices (Hamiltonian and overlap) remains a challenge. To that end, the partition-of-unity finite element method (PUFEM)~\cite{Melenk1996,Babuska1997} ensures locality by modulating the enrichment functions with a set of local polynomials that form a partition-of-unity (i.e., akin to the CFE basis functions) and has been adopted in~\cite{Sukumar2009,Pask2011,Pask2017}. As a result of maintaining the locality of the basis at the same level of the CFE basis, PUFEM simplifies the discrete matrix structure and
load balancing in a parallel computing framework. However, given that each enrichment function in PUFEM are modulated with several local  polynomials, PUFEM entails a large number of additional functions. An alternative approach is to multiply the the enrichment functions with a single smooth cutoff function and has been adopted in ~\cite{kanungo2017large}, in the context of large-scale all-electron calculations. As demonstrated in ~\cite{kanungo2017large}, this particular EFE approach attains a staggering $50-100\times$ speedup over the CFE basis, and a $3-8\times$ speedup over the gaussian basis. While enrichment of the FE basis resulted in impressive improvements in efficiency, such an enrichment is prone to ill-conditioning with increasing refinement of the CFE basis ~\cite{Schweitzer2011,Babuska2012,Pask2017,Albrecht2018, Cai2013}. To elaborate, since, unlike the planewave augmentations, the enrichment functions spatially overlap with the CFE basis functions, they remain susceptible to becoming linearly dependent on the CFE basis. In turn, it affects the robustness and accuracy of the EFE basis, especially while dealing with a refined CFE basis.

The ill-conditioning problem is also present in PUFEM, and several efforts ranging from stabilization~\cite{Babuska2012,Gupta2013} to orthogonalization procedures~\cite{Sillem2015} have been proposed to alleviate the problem. However, these schemes have been designed keeping in view engineering applications (e.g., fracture mechanics, elastostatics) as well as the local structure of the partition-of-unity, and hence, cannot be trivially extended to all-electron DFT calculations involving an EFE basis which does not employ a partition-of-unity. Recently, a combination of flat-top partition-of-unity approach and local partial-orthogonalization~\cite{Schweitzer2011} has been extended to solve the Schr\"odinger equation with a localized potential, attaining an $\mathcal{O}({10^{10}})$ decrease in the condition number over PUFEM~\cite{Albrecht2018}. However, its efficacy for all-electron DFT calculations remains unexplored.

This work presents a robust approach to construct a well-conditioned and local EFE basis for all-electron DFT calculations.
We resolve the ill-conditioning in the EFE basis by introducing an orthogonalized enriched FE (OEFE) basis. To elaborate, we recast the enrichment functions such that they are orthogonal to the underlying CFE basis, while maintaining the locality of the resultant basis. In addition to the orthogonalization of the enrichment functions, in this work, we generalize the enrichment to handle periodic systems. In particular, we employ k-point dependent enrichment functions, so as to afford greater computational efficiency. To efficiently solve for the electrostatic potentials, we use the smeared charge approach proposed in \cite{pask2012linear}. This procedure involves replacing the point nuclear charge by an analytical smeared charge whose corresponding potential can be used to correct for the electrostatic potential. Lastly, as an efficient solution strategy for solving the discrete Kohn-Sham eigenvalue problem, we employ the Chebyshev polynomial based filtering approach~\cite{Zhou2006a,Zhou2006b,Motamarri2013} to compute the subspace spanned by the occupied eigenstates, and then solve the Kohn-Sham eigenvalue problem by projecting the problem onto the Chebyshev-filtered subspace. 

We demonstrate the accuracy and efficiency of the proposed OEFE basis for all-electron DFT calculations, using both unit cell and large scale periodic calculations. First, we study the rate of convergence in the ground-state energy with respect to mesh size by performing $\Gamma$-point calculations on lithium fluoride (LiF) and diamond. Next, we assess the accuracy of the proposed method against the LAPW+lo basis by comparing the k-point converged ground-state energy and band structure on
magnesium sulfide (MgS) and cerium (Ce) unit cells. Additionally, we demonstrate the competence of the OEFE basis for large scale all-electron calculations on  four materials systems: (i) silicon carbide (SiC) divacancy, (ii) NV-diamond, (iii) copper (Cu) monovacancy, and (iv)  silver chloride (AgCl) divacancy, each of increasing supercell sizes. We attain a substantial $130\times$ speedup of the OEFE basis over the CFE basis. 
%\sout{Moreover, the orthogonalized enriched FE competes with the LAPW+lo implementation in the Elk\cite{elkcode} code for modest system sizes.} 
Moreover, the OEFE basis outperforms the LAPW+lo implementation in the Elk\cite{elkcode} code for the moderately sized SiC divacancy and the NV-diamond systems. For systems containing heavier atoms---the Cu monovacancy and the AgCl divacancy systems---the LAPW+lo implementation outperforms the OEFE basis. Notably, with the OEFE basis we are able to perform calculations on large systems, ranging up to 9,980 electrons, using modest computational resources, which are otherwise inaccessible to the LAPW+lo implementation in Elk. Lastly, we study the strong scaling behaviour of the OEFE basis, using a 62 atom SiC divacancy system, and observe an efficiency of 92\% at $22\times$ speedup (192 processors).

% \tcb{\sout{Before proceeding to the rest of the paper, we make a brief note on the partition-of-unity finite element method (PUFEM)~\cite{Melenk1996,Babuska1997} which, similar to the enriched and orthogonalized enriched FE basis, also augments the CFE basis with additional basis functions. Unlike the enriched and orthogonalized enriched FE basis, the enrichment functions in PUFEM are modulated with a set of local polynomials (akin to the CFE basis functions), so as to maintain the same locality as that of the underlying CFE basis. Over the past decade, several efforts~\cite{Sukumar2009,Pask2011, pask2012linear,Pask2017, Albrecht2018} have employed the PUFEM for DFT calculations. In particular, Albrecht et. al.~\cite{Albrecht2018} have combined PUFEM along with flat-top partition-of-unity and local preconditioners~\cite{Schweitzer2011} to address the ill-conditioning problem associated with standard PUFEM as well as cast the discrete Kohn-Sham eigenvalue as a standard eigenvalue problem, as opposed to a generalized eigenvalue problem that is natural to any PUFEM. However, all the PUFEM based efforts at DFT calculations have been restricted to only pseudopotential calculations on small-scale systems. Thus, the efficacy of PUFEM for the all-electron DFT calculations is yet to be ascertained. More importantly, given that each enrichment function in PUFEM is modulated with several local polynomials, PUFEM entails a significantly higher number of additional functions than the enriched and orthogonalized enriched FE basis.}}

The rest of the paper is organized as follows. In Sec.~\ref{sec:form}, we present the real-space formulation for periodic all-electron Kohn-Sham density functional theory calculations employed in this work. The details of the OEFE disretization are presented in Sec.~\ref{sec:oefe}, which is followed by the numerical approach employed in the solution of the discrete Kohn-Sham problem in Sec.~\ref{secscfcf}. In Sec.~\ref{sec:res}, we demonstrate the accuracy, efficiency and parallel scalability of the OEFE basis. Finally, we summarize our findings and present the future scope of this work in Sec.~\ref{sec:summary}.

%%%%%%%%%%%%%%%%%%%%%%%%%%%%%%% FORMULATION %%%%%%%%%%%%%%%%%%%%%%%
% \input{formulation}
\section{\label{sec:form}Formulation of Kohn-Sham DFT}
For periodic systems, the Kohn-Sham eigenvalue problem can be written as
\begin{equation} \label{eqnksdft}
    \left( -\frac{1}{2} \del^{2} + V_{\text{eff}}(\rho,\bR) \right) \psialbk(\bx) = \epsalbk \psialbk(\bx)\,,
\end{equation}
where $\psialbk(\bx)$ and $\epsalbk$ are the Kohn-Sham 
eigenfunctions and eigenvalues, respectively, corresponding to the $\bk$ point in the reciprocal space; the index $\alpha$ runs over all the electrons ($N_e$) in the system; and $\bR = \{\bR_{1},\bR_{2},\ldots,\bR_{N_a}\}$ corresponds to the position of the $N_a$ atoms in the system. The effective Kohn-Sham potential $V_{\text{eff}}(\rho,\bR)$ is constructed using the electron density $\rho(\bx)$. We remark that as the first effort at an OEFE basis for all-electron DFT, we present the formulation in the context of non-relativistic DFT. Nevertheless, the ideas explored can be extended to relativistic DFT (scalar relativistic and spin-orbit coupling) as well. Furthermore, in the current work, we restrict our analysis to spin-independent systems. However, all the ideas discussed subsequently can be generalized, in a straightforward manner, to spin-dependent systems.

The constituents of the effective potential $V_{\text{eff}}(\rho,\bR)$ are given by
\begin{equation}
    V_{\text{eff}}(\rho,\bR) = V_{\text{xc}} \left(\rho\right) +  V_{\text{H}}\left(\rho\right)+ V_{\text{ext}}\left(\bR\right)\,, 
\end{equation}
where $V_{\text{xc}} \left(\rho\right)=\frac{\delta E_{\text{xc}}[\rho]}{\delta \rho(\bx)}$ is the exchange-correlation potential computed as the functional derivative of the exchange-correlation energy $E_{\text{xc}} \left[\rho\right]$ with respect to $\rho$. $V_{\text{xc}} \left(\rho\right)$ is a mean-field potential which accounts for quantum mechanical many-body interactions. In this work, we use the local density approximation (LDA) exchange-correlation functional with Ceperley and Adler constants\cite{perdewzunger,ceperley1980ground}. $V_{\text{H}}$ and $V_{\text{ext}}$ are the Hartree and nuclear potentials, respectively, and are given by
\begin{equation}
    V_{\text{H}}\left(\bx\right) = \underset{\Rthree} \int \frac{\rho(\bx')}{|\bx - \bx'|} d\bx'\,,
\end{equation}
\begin{equation}
    V_{\text{ext}}\left(\bx\right) = - \sum_{J} \frac{Z_J}{|\bx - \bR_{J}|}\,,
\end{equation}
where $Z_J$ is the atomic number of the $J$\textsuperscript{th} nucleus in $\Rthree$. Equivalently, the evaluation of the electrostatic potentials can be recast as a Poisson problem\cite{Pask2005,Suryanarayana2010, Motamarri2012, das2015real}. In this work, we compute the total electrostatic potential which combines both the Hartree and the nuclear potentials. The total electrostatic potential $\phitot(\bx)$ can be evaluated by solving the following Poisson problem
\begin{equation} \label{eqnpoisson}
    -\frac{1}{4\pi} \del^{2} \phitot(\bx) =  b(\bx) + \rho(\bx)\,,
\end{equation}
where $b(\bx)$ is the sum of all nuclear charges.  Conventionally, in an all-electron calculation the nuclear charges are treated as point charges, i.e., 
\begin{equation} \label{eqnpoint}
b(\bx)= - \sum_{I} Z_{I} \tilde{\delta} (\bx - \bR_{I})\,,    
\end{equation}
where $\tilde{\delta}(\bx-\bR_I)$ is the Dirac delta function representing a point nuclear charge at $\bR_I$. Equivalently, as shown in~\cite{pask2012linear}, one could evaluate $\phitot$ using an appropriately scaled smeared charge such that it integrates to the same value as the point charge. To elaborate, we can define a smeared nuclear charge $\bsmear(\bx)$ given as 
\begin{equation} \label{eqnsmeared}
\bsmear(\bx) = - \sum_{I} Z_{I} g(|\bx - \bR_{I}|, r_{c,I})\,,
\end{equation}
where $g(|\bx-\bR_I|,r_{c,I})$ denotes a unit smeared charge which is localized within $|\bx-\bR_I| < r_{c,I}$ and integrates to unity. In this work, we employ the following form for the unit smeared charge~\cite{pask2012linear}
\begin{equation} \label{eqngr}
    g(r,r_c) = \begin{cases}
    \frac{-21 (r-r_c)^3 (6r^2 + 3r r_c + r_c^2)}{5 \pi r_c^8},  & 0\leq r \leq r_c, \\
    0, & r > r_c
    \end {cases}
\end{equation}
The $r_{c,I}$'s are chosen to be the largest possible values that avoid overlap between two neighboring smeared charges. Subsequently, we use $\bsmear$ to compute an auxiliary electrostatic potential $\phiAux(\bx)$ given as 
\begin{equation} \label{eqnpoissonaux}
    -\frac{1}{4\pi} \del^{2} \phiAux(\bx) = \bsmear(\bx) + \rho\left(\bx\right)\,.
\end{equation}
Finally, the total electrostatic potential $\phitot$ is obtained from $\phiAux$ by adding a correction term, and is given as
\begin{equation} \label{eqnpoissontotsmear}
\begin{split}
      \phitot(\bx) =& \phiAux(\bx) +\\
      &\sum_{I}   \left(V_{\text{N},I}\left(|\bx - \bR_I|\right)-V_{\text{N},I}^{\text{smear}}\left(|\bx - \bR_I|,r_{c,I}\right)\right)\,,
\end{split}
\end{equation}
where the second term is the correction term comprising of the sum of the difference between the exact nuclear potential ($V_{\text{N},I}$) and the smeared nuclear potential ($V_{\text{N},I}^{\text{smear}}$), both corresponding to the $I$\textsuperscript{th} nucleus. The exact and smeared nuclear potentials for the $I$\textsuperscript{th} nucleus are given by
\begin{equation} \label{eqnnuc}
    V_{\text{N},I}(r) = -\frac{Z_I}{r}\,,
\end{equation}
\begin{equation} \label{eqnnucsmear}
    V_{\text{N},I}^{\text{smear}}(r,r_{c,I}) = -Z_I v_g(r,r_{c,I})\,,
\end{equation}
where $v_g(r,r_c)$ is the potential corresponding to the $g(r,r_c)$ and is given by
\begin{equation} \label{eqnvg}
    v_g(r,r_c) = \begin{cases}
    \frac{9r^7 - 30r^6r_c + 28r^5r_c^2 - 14r^2r_c^5 +12r_c^7}{5r_c^8}, &0\leq r \leq r_c \\
    \frac{1}{r}, &r > r_c\,.
    \end{cases}
\end{equation}
At this juncture, we note that in a sufficiently refined FE basis, as is typically warranted in a CFE basis based all-electron calculation, both the point and smeared charge approaches provide comparable accuracy. However, while using a coarse FE basis, as is the case while employing an EFE basis, the smeared charge approach fares better in terms of computational efficiency. Thus, for the remaining of the paper, we restrict our discussion regarding the electrostatic potential to the smeared charges. 

Returning to Eq.~\ref{eqnksdft}, we invoke the Bloch theorem~\cite{Ashcroft1976} to write $\psialbk\left(\bx\right)$ in terms of the Kohn-Sham periodic function, $\ualbk\left(\bx\right)$---a quantity which respects the periodicity of the crystal---and is given by
\begin{equation}
    \psialbk \left(\bx\right) = \exp \left(i\bk\cdot\bx \right) \ualbk  \left(\bx\right)\,.
\end{equation}
Using the above relation, Eq. \ref{eqnksdft} becomes
\begin{equation} \label{eqnksdftu}
\begin{split}
    &\left( -\frac{1}{2} \left(\del^2 + 2i\bk\cdot\del -|\bk|^2\right) + V_{\text{eff}}(\rho,\bR) \right) \ualbk(\bx) =\\
    & \quad \epsalbk \ualbk(\bx)\,.
    \end{split}
\end{equation}
The electron charge density is computed in terms of $\ualbk$ as follows
\begin{equation} \label{eqnrho}
    \rho\left(\bx\right) = 2 \sum_{\bk} w_{\bk} \sum_{\alpha} f\left( \epsalbk,\mu \right) |\ualbk\left(\bx\right)|^2\,,
\end{equation}
where $f\left( \epsalbk , \mu \right)$ is the fractional occupancy of the eigenstate with eigenvalue $\epsalbk$, and $\mu$ denotes the Fermi level. The sum over $\bk$ is a discrete sum over grid points lying in the Brillouin zone, and $w_{\bk}$ is the associated weight. Typically, the k-point grid is chosen based on the Monkhorst-Pack (MP) scheme\cite{monkhorst1976special}. We use the Fermi-Dirac distribution for the fractional occupancy, given by 
\begin{equation} \label{eqnfd}
    f\left( \epsilon , \mu \right) = \frac{1}{1+\exp{\left( \frac{\epsilon - \mu }{k_B T}\right)}}\,,
\end{equation}
where $T$ is the smearing temperature and $k_B$ is the Boltzmann constant. The Fermi level $\mu$ is determined by the constraint on the number of electrons $N_e$ in the simulation domain $\Omega$, and is given by
\begin{equation} \label{eqnne}
    \int_{\Omega} \rho\left(\bx\right) = 2 \sum_{\bk} w_{\bk} \sum_{\alpha} f\left( \epsalbk, \mu \right) = N_e\,.
\end{equation}
We remark that by exploiting the symmetry of the crystal\cite{dresselhaus2007group,togo2018textttspglib}, we can achieve a reduction\cite{richardMartin} in the number of $\bk$-points of the MP grid for which $\ualbk$ needs to be computed. To elaborate, if $\bk_1$ and $\bk_2$ belong to an MP grid and $\bk_2 = \hat{\bR} \bk_1 $, where $\hat{\bR}$ is a point group operation of the crystal, we have
\begin{align}
u_{\alpha,\bk_1}\left(\bx\right) = u_{\alpha,\bk_2}(\hat{\bR} \bx + \hat{f} ),\quad \text{and}  && \epsilon_{\alpha,\bk_1} = \epsilon_{\alpha,\bk_2}\,,
\end{align} 
where $\hat{f}$ is a fractional translation~\cite{richardMartin} corresponding to $\hat{\bR}$.

Finally, upon solving Eq.~\ref{eqnksdftu} and Eq.~\ref{eqnrho} self-consistently, the ground-state energy of the system is given by  
\begin{equation}
    E_{\text{tot}} = E_{\text{band}} + E_{\text{xc}} + E_{\text{elec}} - \intomega \rho V_{\text{xc}} \dx - \intomega \rho \phi_{\text{tot}} \dx\,.
\end{equation}
In the above equation, $E_{\text{band}}$ is the band energy given by 
\begin{equation}
    E_{\text{band}} = 2 \sum_{\bk} w_{\bk} \sum_{\alpha} f\left( \epsalbk , \mu \right) \epsalbk\,.
\end{equation}
$E_{\text{elec}}$ is the electrostatic energy given by~\cite{pask2012linear}
\begin{equation}
\begin{split}
    E_{\text{elec}} &= \intomega \frac{1}{2} (\bsmear(\bx) + \rho(\bx) )\phiAux(\bx) d\bx  + \\
    &\sum_I \intomegai \rho(\bx) \left(V_{\text{N},I}\left(|\bx - \bR_I|\right) - V_{\text{N},I}^{\text{smear}}\left(|\bx - \bR_I|,r_{c,I}\right) \right) + \\
    &\sum_I \frac{1}{2} Z_I^2 (I_g(r_{c,I}) - v_g(0,r_{c,I}))\,,
\end{split}
\end{equation}
where $\Omega_I$ denotes a sphere of radius $r_{c,I}$ centered at $\bR_I$ (i.e., the compact support of $(V_{\text{N},I}-V^{\text{smear}}_{\text{N},I})$), and $I_g(r_c) = 10976/(17875r_c)$\,.

%%%%%%%%%%%%%%%%%%%%%%% OEFEM %%%%%%%%%%%%%%%%%%%%%%%%%%%%%%%%%
%\input{oefem}
\section{Orthogonalized enriched finite element (OEFE) method} \label{sec:oefe}

We now present the details of the OEFE discretization proposed in this work. To begin with, the EFE discretization~\cite{kanungo2017large} augments the CFE basis (a continuous localized piecewise polynomial basis~\cite{hughes2012finite,bathe2006finite}) with atom-centered numerical basis---termed as enrichment functions. The key idea here is to account for the sharp variations in the orbitals and the electrostatic potentials close to nuclei, largely,
through the enrichment functions, and thereby eliminate the need for a refined classical finite element mesh close to the nuclei. Although it offers an efficient basis for all-electron calculations, it can result in an ill-conditioned basis. To elaborate, the enrichment functions remain prone to being linearly dependent on the CFE basis, especially while using refined finite element meshes, thus potentially affecting the accuracy and robustness of the EFE basis. While one can control the ill-conditioning by decreasing the compact support of the enrichment functions through a smooth cutoff function (as adopted in Ref.~\cite{kanungo2017large}), it results in three issues: (i) the decrease in the compact support leads to a deterioration of the enrichment functions, in terms of capturing the electronic fields, and hence, results in the use of higher number of CFE basis functions to compensate; (ii) the improvement to the conditioning through this approach is both limited and marginal (as will be demonstrated in Sec.~\ref{sec:cond}); (iii) a reasonable choice of truncating the enrichment functions becomes dependent on the underlying classical FE mesh, thereby affecting the ease and robustness of generating a suitable EFE basis. To this end, we alleviate the problem of ill-conditioning in the EFE basis by formulating an OEFE basis.

\subsection{Orthogonalized enriched finite element discretization}
The OEFE discretization of the Kohn-Sham periodic functions, $(\ualbk^h(\bx))$, is given by
\begin{equation} \label{eqnUOEFE}
    \ualbk^h(\bx)=\underbrace{\sum_{i=1}^{n_h}\shp{C}{i}u_{\alpha,\bk,i}^C}_{\text{Classical}} + \underbrace{\sum_{I=1}^{N_a}\sum_{j=1}^{n_I}\shp{O,u_{\bk}}{j,I}u_{\alpha,\bk,j,I}^{O}}_{\text{Orthogonalized Enriched}}\,.
\end{equation}
In the above equation, the superscript $h$ indicates a discrete field, and the superscript $C$ and $O$ are used to distinguish the classical and the orthogonalized enriched components, respectively. $\shp{C}{i}$ denotes the $i\textsuperscript{th}$ CFE basis function, and $u_{\alpha,\bk,i}^C$ denotes the expansion coefficient of $\shp{C}{i}$ for $\ualbk$. Similarly, $\shp{O,u_{\bk}}{j,I}$ denotes the k-point dependent orthogonalized enrichment function for $\ualbk~(\forall \alpha)$. The index $I$ runs over all the atoms ($N_a$) in the system, and the index $j$ runs over all the atomic Kohn-Sham orbitals ($n_I$) we include for the atom $I$. In other words, the $I$\textsuperscript{th} atom, situated at $\bR_I$, contributes $n_I$ enrichment functions, each centered around $\bR_I$. $u_{\alpha,\bk,j,I}^O$ represents the expansion coefficient of $\shp{O,u_{\bk}}{j,I}$ corresponding to $\ualbk$.  
%Also, we make a note that, for periodic systems, $\bR_I$ represents not just the nuclear position in the simulation domain, it also refers to the images of the  $I$\textsuperscript{th} nucleus in the nearest translated cells. To evaluate the value of the enrichment at a point $\bx$, we simply pick an  $\bR_I$ which yields the smallest $|\bx-\bR_I|$. This ensures that the enrichment functions wrap around the periodic faces making the basis for $\ualbk^h(\bx)$ periodic.

Turning to the form of the orthogonalized enrichment function, $\shp{O,u_{\bk}}{j,I}$, we split it into two parts, given as
\begin{equation} \label{eqnUOEFESplit}
    \shp{O,u_{\bk}}{j,I} = \shp{A,u_{\bk}}{j,I} - \shp{B,u_{\bk}}{j,I}\,.
\end{equation}
In the above equation, $\shp{A,u_{\bk}}{j,I}$ is the atomic part that encapsulates the single atom Kohn-Sham orbital information. On the other hand, $\shp{B,u_{\bk}}{j,I}$ denotes the component of $\shp{A,u_{\bk}}{j,I}$ along the CFE basis which, when subtracted from $\shp{A,u_{\bk}}{j,I}$, guarantees the orthogonality of $\shp{O,u_{\bk}}{j,I}$ with respect to the CFE basis $\{\shp{C}{i}\}$.

%Next, we comment on the form of $\shp{A,u_{\bk}}{j,I}$. Sukumar and Pask\cite{Sukumar2009} have previously demonstrated various forms of enrichment functions for periodic problems. 
We note that $\shp{A,u_{\bk}}{j,I}$ needs to be both periodic as well as incorporate k-point dependence. To that end, we
choose $\shp{A,u_{\bk}}{j,I}$ to be
\begin{equation}\label{eqnUOEFEAtomic}
    \shp{A,u_{\bk}}{j,I} = e^{-i\bk.(\bx-\bRIx)} \tilde{\psi}_{j,I}(\bx,\bRIx)\,.
\end{equation}
In the above equation, $\bRIx$ denotes position of the nearest image of the atom at $\bR_I$ to $\bx$. In other words, for a given point $\bx$ and nucleus $I$, $\bRIx$ is the position chosen from the set consisting of $\bR_I$ and its periodic images such that it yields the smallest distance from $\bx$. This ensures periodicity of the enrichment functions by wrapping them around the periodic boundaries. The function $\tilde{\psi}_{j,I}(\bx,\bRIx)$ is a truncated Kohn-Sham orbital of the isolated atom of the atom type located at $\bRIx$ (i.e., of the atom of the $I^\text{th}$ nucleus). $\tilde{\psi}_{j,I}(\bx,\bRIx)$ is given by
\begin{equation} \label{eqnKSAtomicTrunc}
    \tilde{\psi}_{j,I}(\bx,\bRIx) = \psi_{nlm,I}(|\bx-\bRIx|,\beta_{\bRIx},\gamma_{\bRIx})h(|\bx-\bRIx|,r_0,t)\,,
\end{equation}
where $\psi_{nlm,I}$ is an atomic Kohn-Sham orbital indexed by the principal quantum number $n$, azimuthal quantum number $l$, and magnetic quantum number $m$, for an isolated atom of the atom type of the $I^\text{th}$ nucleus, defined in spherical coordinates. We maintain an appropriate correspondence between $j$ and $(n,l,m)$. $\beta_{\bRIx}$ and $\gamma_{\bRIx}$ are the polar and azimuhthal angles, respectively, with respect to a shifted origin at $\bRIx$. Typically, we include all the $\psi_{nlm,I}$ with non-zero fractional occupancy as enrichment functions. The function $h(r,r_0,t)$ is a smooth cutoff function, parameterized by a cutoff radius $r_0$ and smoothness factor $t$, and has the following properties,
\begin{align} \label{eqnCutoffProperties}
h(r,r_0,t) = 
\begin{cases} 
1 & 0 \leq r < r_0, \\
    0 \leq  h < 1  & r_0 < r \leq r_0 + \frac{r_0}{t}\,,\\
    0 &   r > r_0 + \frac{r_0}{t}\,.
\end{cases}
\end{align}
We remark that $h(r,r_0,t)$ offers two vital functions: (i) it avoids spurious self interaction of the enrichment functions, especially for periodic problems with small lattice constants; and (ii) it renders locality to the basis, and hence, is crucial to the parallel efficiency of its implementation. We refer to~\cite{kanungo2017large} for a detailed discussion on the form of $\psi_{nlm,I}$ and $h(r,r_0,t)$ as well as the choices for $r_0$ and $t$. In particular, we localize the enrichment functions within $2.5-3.5$ a.u.\ from its corresponding nucleus or to the maximum extent feasible to avoid self-interaction, whichever is smaller. In this work, we use the radial Kohn-Sham solver code  \textit{dftatom}\cite{CERTIK20131777} to precompute $\psi_{nlm,I}$.

At this stage, for simplicity of notation, we combine the $\{j,I\}$ indices in $\shp{O,u_{\bk}}{j,I}$, $\shp{A,u_{\bk}}{j,I}$, $\shp{B,u_{\bk}}{j,I}$  into a single index $\nu$. Further, we define
$n_O^{u}=\sum_{I=1}^{N_a} n_I$ to denote the total number of orthogonalized enrichment functions used for discretizing $\ualbk$.

We now turn to $\shp{B,u_{\bk}}{\nu}$---the orthogonalizing part of $\shp{O,u_{\bk}}{\nu}$ (Eq.~\ref{eqnUOEFESplit}). Given that $\shp{B,u_{\bk}}{\nu}$ represents the component of $\shp{A,u_{\bk}}{\nu}$ along the CFE basis, we define it as
\begin{equation} \label{eqnUOEFEOrtho}
    \shp{B,u_{\bk}}{\nu} = \sum_{l=1}^{n_h} c_{\nu,l}^{\bk}\shp{C}{l}\,,
\end{equation}
where the coefficients $c_{\nu,l}^{\bk}$ are to be obtained using the orthogonality condition,
\begin{equation} \label{eqnOrthoCondition}
    \int_{\Omega} \shp{O,u_{\bk}}{\nu} \shp{C}{j}\dx = 0\,, \quad  j=1,2,\ldots,n_h\,.
\end{equation}
From Eq.\ref{eqnUOEFESplit} and Eq.\ref{eqnOrthoCondition}, we get
\begin{equation} \label{eqnorthC}
    \bM^{\textbf{cc}}\bc_{\nu}^{\bk}=\bd_{\nu}^{\bk}\,.
\end{equation}
In the above equation, $\bM^{\textbf{cc}}$ is the overlap matrix of the CFE basis, given by 
\begin{equation} \label{eqnMcc}
    M^{cc}_{jl}=\int_{\Omega}\shp{C}{j}\shp{C}{l}\dx\,,
\end{equation}
and $\bc_{\nu}^{\bk}$ is the vector containing the coefficients $c_{\nu,l}^{\bk}$, and the vector $\bd_{\nu}^{\bk}$ is defined as
\begin{equation}
    d_{\nu,j}^{\bk}=\int_{\Omega}\shp{A,u_{\bk}}{\nu}\shp{C}{j}\dx\,.
\end{equation}
We further simplify the evaluation of $\bc_{\nu}^{\bk}$ by employing a combination of spectral finite-elements and Gauss-Lobatto-Legendre (GLL) quadrature rule, which renders $\bM^{\textbf{cc}}$ diagonal. Typically, the CFE basis functions are the Lagrange polynomials generated using equidistant nodes in the finite element. In spectral finite elements, however, the Lagrange polynomials are generated using the Gauss-Lobatto-Legendre (GLL) node distribution\cite{boyd2001chebyshev}. Thus, the use of spectral finite elements along with the GLL quadrature rule make the nodal points and the quadrature points coincident, resulting in the CFE overlap matrix ($\bM^{\textbf{cc}}$) being diagonal. We refer to~\cite{Motamarri2013} for an elaborate discussion on spectral finite elements. 
Thus, the use of spectral finite-elements and GLL quadrature simplifies the evaluation of $\bc_{\nu}^{\bk}$ in Eq.~\ref{eqnorthC} to
\begin{equation} \label{eqnorthCGLL}
    c_{\nu,l}^{\bk} = \frac{\int_{\Omega} \shp{A,u_{\bk}}{\nu} \shp{C}{l}\dx}{\int_{\Omega,\text{GLL}} \shp{C}{l} \shp{C}{l}\dx}\,.
\end{equation}
We remark that the function $\shp{A,u_{\bk}}{\nu}$ has a compact support governed by the smooth cutoff function $h(r,r_0,t)$ (Eq.~\ref{eqnCutoffProperties}). Thus, $c_{\nu,l}^{\bk}$ is non-zero for only those $l$ indices for which $\shp{C}{l}$ has an overlap with the compact support of $\shp{A,u_{\bk}}{\nu}$. In other words, $\shp{O,u_{\bk}}{\nu}$ retains the locality of $\shp{A,u_{\bk}}{\nu}$.

Having constructed the OEFE basis for $\ualbk$, the discrete Kohn-Sham eigenvalue problem corresponding to a k-point $\bk$ is obtained by using Eq.~\ref{eqnUOEFE} in Eq.~\ref{eqnksdftu} and is given by
\begin{equation} \label{eqnKSEigGenDiscrete}
    \bH_{\bk}^O \textbf{u}_{\alpha,\bk}^O=\epsilon_{\alpha,\bk}^O \bM_{\bk}^O \textbf{u}_{\alpha,\bk}^O\,,
\end{equation}
where $\textbf{u}_{\alpha,\bk}^O$ is the eigenvector containing the coefficients $u_{\alpha,\bk,j}^C$ and $u_{\alpha,\bk,\nu}^O$ (see Eq.~\ref{eqnUOEFE}), and $\epsilon_{\alpha,\bk}^O$ is its corresponding eigenvalue. $\bH_{\bk}^O$ represents the discrete Hamiltonian matrix and is given by,
\begin{equation} \label{eqnHOEFEelelements}
\begin{split}
    H_{\bk,mn}^O &= \frac{1}{2} \intomega \del N_m^{\dagger}(\bx) \cdot \del N_n(\bx) dx  \\
     & - \intomega i \bk \cdot \left( N_m^{\dagger}(\bx) \del N_n(\bx)  \right)  \dx   \\
    & + \intomega \left( \frac{|\bk|^2}{2} + V^h_{\text{eff}}\left(\bx,\bR\right)  \right) N_m^{\dagger}(\bx)  N_n(\bx) \dx\,,
    \end{split}
\end{equation}
where the superscript $\dagger$ denotes the complex conjugate, and the functions $N_m(\bx)$, $N_n(\bx)$ are generic representations for $\shp{C}{j}$ and $\shp{O,u_{\bk}}{\nu}$. The matrix $\bM_{\bk}^O$ is the overlap matrix and has the following block-diagonal structure, owing to the orthogonality condition of Eq.~\ref{eqnOrthoCondition},
\begin{equation} \label{eqnMOEFE}
\bM_{\bk}^O=\left[
\begin{array}{c|c}
\bM^{\textbf{cc}}  & 0 \\ \hline
 0 &  \bM_{\bk}^{\textbf{oo}}
\end{array}\right]\,.
\end{equation}
In the above equation, $\bM^{\textbf{cc}}$ block contains the overlap between two CFE basis functions (as defined in Eq.~\ref{eqnMcc}), and the $\bM_{\bk}^{\textbf{oo}}$ block contains the overlap between two orthogonalized enrichment functions, i.e.,
\begin{equation} \label{eqnMoo}
    M^{oo}_{\nu\mu}=\int_{\Omega}\shp{O,u_{\bk}}{\nu}\shp{O,u_{\bk}}{\mu}\dx\,.
\end{equation}

We note that Eq. \ref{eqnKSEigGenDiscrete} is a generalized eigenvalue problem. However, we can invert $\bM^O_{\bk}$ to transform it into a standard eigenvalue problem, given by 
\begin{equation} \label{eqnKSEignStandDiscrete}
    (\bM_{\bk}^{O})^{-1} \bH_{\bk}^O \textbf{u}^O_{\alpha,\bk} = \epsilon_{\alpha,\bk}^O \textbf{u}^O_{\alpha,\bk}\,.
\end{equation}
We emphasize that the above transformation to a standard eigenvalue problem is crucial to our use of the Chebyshev polynomial based  filtering technique as an efficient eigensolver (see Sec. \ref{secscfcf}). Naturally, the transformation to the standard eigenvalue warrants efficient means to invert $\bM_{\bk}^{O}$. The inverse of $\bM_{\bk}^O$ also has a block diagonal form, given by
\begin{equation} \label{eqnMOEFEInv}
(\bM_{\bk}^O)^{-1}=\left[
\begin{array}{c|c}
(\bM^{\textbf{cc}})^{-1} & 0 \\ \hline
0  &   \left( \bM_{\bk}^{\textbf{oo}} \right)^{-1}
\end{array}\right]\,.
\end{equation}
As discussed earlier, the evaluation of $(\bM^{\textbf{cc}})^{-1}$ is trivial, given that $\bM^{\textbf{cc}}$ is rendered diagonal through the combined use of spectral finite elements and GLL quadrature. The $\left(\bM^{\textbf{oo}} \right)^{-1}$ block, being a small dense matrix of size $n_O^u\times n_O^u$, is evaluated through direct solvers. 

Finally, we turn to the OEFE discretization of the auxiliary electrostatic potential $\phi_{\text{aux}}^h$ (Eq.~\ref{eqnpoissonaux}), given as  
\begin{equation} \label{eqnPhiOEFE}
    \phi_{\text{aux}}^h(\bx)=\underbrace{\sum_{j=1}^{n_h}\shp{C}{j}\phi_{j}^C}_{\text{Classical}} + \underbrace{\sum_{I=1}^{N_a}\shp{O,\phi}{I}\phi_{I}^O}_{\text{Orthogonalized Enriched}}\,,
\end{equation}
where the supersript $C$ and $O$ denote the classical and orthogonalized enriched components, respectively. As with the discretization of $\ualbk$ (Eq.~\ref{eqnUOEFE}), $\shp{C}{j}$ denotes the $j\textsuperscript{th}$ CFE basis function and $\phi_j^C$ denotes its corresponding coefficient. Similarly, $\shp{O,\phi}{I}$ is the $I\textsuperscript{th}$ orthogonalized enrichment function 
with a corresponding coefficient $\phi_I^C$.  
Similar to $\ualbk$, the enrichment function for $\phiAux$ is also split into two parts, given by
\begin{equation} \label{eqnOrthoEnrichedPhiAux}
    \shp{O,\phi}{I} = \shp{A,\phi}{I}  -
    \shp{B,\phi}{I}\,,
\end{equation}
where $\shp{A,\phi}{I}$ and $\shp{B,\phi}{I}$ are the atomic and orthogonalizing parts, respectively. The atomic part, $\shp{A,\phi}{I}$, is expressed as
\begin{equation} \label{eqnPhiAuxAtomic}
    \shp{A,\phi}{I} =  \phi_{\text{aux},I}(\bx)h(|\bx-\bRIx|,r_0,t)\,,
\end{equation}
where $\bRIx$ is same as that defined in Eq.~\ref{eqnUOEFEAtomic} and $h(r,r_0,t)$ is the smooth cutoff function defined in Eq.~\ref{eqnCutoffProperties}.  $\phi_{\text{aux},I}$ is the atomic auxiliary potential given as
\begin{equation} \label{eqnEnrichedPhiAux}
    \phi_{\text{aux},I} (\bx) = V_{\text{H},I}(|\bx - \bRIx|) + V_{\text{N},I}^{\text{smear}}(|\bx - \bRIx|,r_{c,I})\,,
\end{equation}
where $V_{\text{H},I}(r)$ denotes the radial Hartree potential of an isolated atom of the same type as located at $\bR_I$, and  $V_{\text{N},I}^{\text{smear}}(r,r_{c,I})$ is the smeared nuclear potential defined in Eq.~\ref{eqnnucsmear}. The orthognalizing part, $\shp{B,\phi}{I}$, of $\shp{O,\phi}{I}$ is evaluated similar to $\shp{B,u_{\bk}}{\nu}$ (Eq.~\ref{eqnUOEFEOrtho}). That is, $\shp{B,\phi}{I}$ is defined as a linear combination of $\{\shp{C}{j}\}$ which guarantees the orthogonality of $\shp{O,\phi}{I}$ with respect to $\{\shp{C}{j}\}$.    

Finally, employing the OEFE discretization of $\phiAux$ in Eq.~\ref{eqnpoissonaux} results in the following discrete Poisson problem 
\begin{equation} \label{eqnpoissX}
     \bA^O \boldsymbol{\phi}^O = \vecc^O\,,
\end{equation}
where $\boldsymbol{\phi}^O$ is the vector containing coefficients $\phi_j^C$ and $\phi_I^O$. $\bA^O$ is the Laplace operator discretized in the OEFE basis for $\phiAux$ and is given by
\begin{equation} \label{eqnDiscLaplace}
    A_{mn}^E = \frac{1}{4\pi} \intomega \del N_m(\bx) \cdot \del N_n(\bx) \dx\,,  
\end{equation}
where $N_m(\bx)$ and $N_n(\bx)$ are generic representations for $\shp{C}{j}$ and $\shp{O,\phi}{I}$. The vector $\mathbf{c}^O$ is the forcing vector, given by
\begin{equation}\label{eqnDiscForceVec}
    c_m^O = \intomega (\bsmear(\bx) + \rho\left(\bx\right) ) N_m(\bx) \dx\,.
\end{equation}

\subsection{Adaptive quadrature}\label{sec:adaptiveQuadrature}
The enrichment functions, $\shp{O,u_{\bk}}{\nu}$ and $\shp{O,\phi}{I}$, are characterized by sharp gradients or oscillations near the nuclei. As a result, an accurate evaluation of the integrals involving the orthogonalized enrichment functions warrants a high quadrature density near the nuclei. However, a uniformly high quadrature density throughout the domain would be inefficient, given that the enrichment functions have a small compact support. To this end, we strike a balance of accuracy and efficiency by using an adaptive quadrature. The key idea is to adopt a divide and conquer strategy in constructing the quadrature grid, based on certain trial integrals~\cite{Berntsen1991,Pieper1999}. In the context of the EFE basis this entails recursively refining each finite element of spatial extent $\Omega_e$ until a set of trial integrals, involving the enrichment functions, attain convergence~\cite{Pask2011, Mousavi2012, Pask2017,kanungo2017large}. We refer to~\cite{Mousavi2012} for an enrichment function based adaptive
quadrature in the context of an EFE basis for DFT and to ~\cite{kanungo2017large} for
the specific details of the adaptive quadrature strategy employed in this
work.

%%%%%%%%%%%%%%%%%%%%%%%%%%%%% SCFCHEBY %%%%%%%%%%%%%%%%%%%%%%%%%%
%\input{scfcheby}
\section{Self Consistent Field Iteration and Chebyshev Filtering} \label{secscfcf}
The Kohn-Sham eigenvalue problem in Eq.~\ref{eqnksdft} is a non-linear eigenvalue problem as the Kohn-Sham Hamiltonian depends on the electron density, which in turn depends on the Kohn-Sham eigenfunctions that are solutions of the eigenvalue problem. Thus, the Kohn-Sham equations, which can be viewed as a fixed point problem, are solved using a self-consistent field (SCF) iteration. The SCF iteration involves using a starting guess density, $\rho_{\text{in}}$, that is used to construct
$V_{\text{eff}}$. Subsequently, the eigenstates $\{\epsalbk,\ualbk\}$ are evaluated, and are, in turn, used to evaluate the output density $\rho_{\text{out}}$. If $||\rho_{\text{in}}(\bx) - \rho_{\text{out}}(\bx)||$ (in an appropriately chosen norm) drops below a tolerance, we declare convergence and compute the ground-state properties corresponding to $\rho_{\text{out}}(\bx)$. Otherwise, $\rho_{\text{in}}$ is updated by mixing~\cite{Anderson1965,Broyden1965,Eyert1996,Kudin2002} $\rho_{\text{in}}$ and $\rho_{\text{out}}$ from previous iterations, and the iteration is continued until convergence in the density.

Computationally, the discrete eigenvalue problem shown in Eq.~\ref{eqnKSEignStandDiscrete} is the most expensive step in each SCF iteration. The dimension of this problem ranges between $\mathcal{O}(10^3)$ to $\mathcal{O}(10^6)$ per atom depending on the species of the atom and our choice of discretization (OEFE basis or CFE basis). Fortunately, we only need to compute the occupied states, i.e. a fraction of the eigenstates at the lower end of the spectrum given
by $N_Y = N_e/2 + N_b$. Here, $N_b$ is a small buffer maintained to capture states with fractional occupancy due to Fermi-Dirac smearing. We compute these eigenstates by using the Chebyshev filtering technique~\cite{Zhou2006a,Zhou2006b,Motamarri2013}. The advantages of this technique over other Krylov subspace methods like the Jacobi-Davidson and Krylov-Schur, in the context of finite element discretization, has been previously demonstrated~\cite{Motamarri2013}. The Chebyshev filtering technique
involves approximating the occupied eigenspace from an initial set of vectors $\mathbf{Y}$ of dimension $N_Y$. A Chebyshev polynomial of degree $m$, $p_m(x)$, exhibits two salient properties: (i) it grows rapidly outside $[-1,1]$, and (ii) $|p_m(x)| \leq 1$ for $x \in [-1,1]$. Thus, given a set of vectors $\mathbf{Y}$, the Chebyshev filtering provides a recipe to construct a new set of vectors, $\widetilde{\mathbf{Y}}$, which spans a subspace that is a close approximation to the occupied eigenspace of interest. The Chebyshev filtered vectors are given by
\begin{equation}
    \widetilde{\mathbf{Y}}=p_m\left(\widetilde{\bH}_{\bk}^O\right)\mathbf{Y}\,,
\end{equation}
where $\widetilde{\bH}_{\bk}^O$ denotes a linear transformation of ${(\bM_{\bk}^O)}^{-1}\bH_{\bk}^O$ such that the unoccupied eigenspectrum of ${(\bM_{\bk}^O)}^{-1}\bH_{\bk}^O$ is mapped to $[-1,1]$ and the occupied spectrum is mapped to $(-\infty,-1)$. In other words, $p_m\left(\widetilde{\bH}_{\bk}^O\right)$ dampens the components of the vectors in $\mathbf{Y}$ that lie along the unoccupied eigenspace and amplifies those lying along the occupied eigenspace. For the purpose of numerical conditioning, we orthonormalize $\widetilde{\mathbf{Y}}$ to produce a set of
orthonormal vectors $\bQ$. Subsequently, we simplify the large eigenvalue problem in Eq.~\ref{eqnKSEignStandDiscrete} by performing a Galerkin projection onto $\bQ$ and solving the following reduced generalized eigenvalue problem   
\begin{equation} \label{eqnProjKS}
\bH_{\bk}^Q \widetilde{\mathbf{u}}_{\alpha,\bk} = \epsalbk^Q \bM_{\bk}^Q \widetilde{\mathbf{u}}_{\alpha,\bk}\,,
\end{equation}
where $\bH_{\bk}^Q = \bQ^{\dagger} \bH_{\bk}^O \bQ$, $\bM_{\bk}^Q = \bQ^{\dagger} \bM_{\bk}^O \bQ$, and $\widetilde{\mathbf{u}}_{\alpha,\bk}$ denotes the eigenvector represented in the Chebyshev filtered subspace. Having solved the above reduced eigenvalue problem, we rotate the eigenvectors to obtain the eigenvectors in the original space, given as:  $\mathbf{u}_{\alpha,\bk}^O=\bQ\widetilde{\mathbf{u}}_{\alpha,\bk}$. Lastly, the set of vectors $\bY$ is updated to $\bQ$ for the next SCF iteration. We note that although
the above procedure is shown in the context of the OEFE basis, it holds even for the CFE basis~\cite{Motamarri2013}. The cost of Chebyshev filtering is determined by the degree of Chebyshev polynomial $m$ required to attain chemical accuracy, which, in turn, is governed by the largest eigenvalue of ${(\bM_{\bk}^O)}^{-1} \bH_{\bk}^O$. The largest eigenvalue increases as the finite element mesh is progressively refined. In other words, the required Chebyshev
polynomial degree, $m$, increases with mesh refinement. In an all-electron calculation, to capture the core states, the rapidly oscillating valence states, and the sharp electrostatic potential near the nucleus, the CFE discretization requires a highly refined mesh in the region. As a consequence, it suffers from the dual disadvantage of requiring large number of degrees of freedom (DoF) as well as a high Chebyshev polynomial degree, $\mathcal{O}(10^3)$, to compute the occupied
eigenspace. In contrast, the OEFE discretization requires a much coarser mesh, as the oscillatory orbitals and the sharp electrostatic potentials near the nuclei are, largely, embedded in the enrichment functions. As a result, the OEFE basis accrues two benefits---a substantial reduction in both the DoF (to obtain chemical accuracy) and the Chebyshev polynomial degree. We illustrate these advantages in the next section.

%%%%%%%%%%%%%%%%%%%%%%%%%%%%%%%%%% RESULTS %%%%%%%%%%%%%%%%%%%%%%%%%%%
%\input{results/results}
\section{\label{sec:res}Results and Discussion}
In this section, we present the numerical results that demonstrate the accuracy and efficacy of the proposed OEFE basis based all-electron calculations. To begin with, we provide a comparative study of the conditioning of the CFE, the EFE, and the OEFE basis, which forms the basis of our adoption of the OEFE basis. Next, we provide the rate of convergence of the ground-state energy with respect to mesh refinement for two benchmark
systems---an 8-atom carbon cubic diamond lattice and an 8-atom halite lithium fluoride (LiF). For the purpose of demonstrating the accuracy of our OEFE basis, we compare the ground-state energies and band-structure against those obtained from LAPW+lo based calculations, for two unit cell systems---8-atom halite magnesium sulfide (MgS) and 4-atom cerium (Ce) face centered cubic (FCC) unit cell. We demonstrate the performance of the OEFE basis for large-scale
all-electron calculations by considering four sets of supercells of varying sizes: (i) divacancy in silicon carbide (SiC), with the largest system containing 9,980 electrons; (ii) NV-diamond; (iii) monovacancy in copper (Cu); and (iv) divacancy in silver chloride (AgCl). For the supercell calculations, we also provide, wherever possible, an accuracy and efficiency comparison against CFE and LAPW+lo basis. All our LAPW+lo calculations are performed using the Elk code~\cite{elkcode}. We note that Elk, by default, employs a relativistic calculation. Thus, in order to conduct a non-relativistic calculation, we suppressed the relativistic effects by scaling the speed of light by a factor 1000 (i.e., by setting the \emph{solscf}(speed of light scaling factor) parameter in Elk to 1000.0). We use an n-stage Anderson mixing~\cite{Anderson1965} for density mixing in all our OEFE and CFE calculations. We use a Fermi-Dirac smearing at 500K in all our calculations to evaluate the occupation number of the Kohn-Sham orbitals. For all calculations involving the OEFE basis, excepting the conditioning studies (Sec.~\ref{sec:cond}) and the supercell calculations (Sec.~\ref{sec:supercellcalcsnew}), we use a uniform FE mesh. For both these calculations, we use an unstructured FE mesh that is refined closer to the nuclei and coarser away from the nuclei. Lastly, we present both the computational complexity (scaling with number of electrons) and strong scaling (scaling with number of processors) of our OEFE implementation. 

\subsection{\label{sec:cond} Conditioning of the basis}
\begin{figure}[htbp]
    \centering
    \includegraphics[scale=0.9]{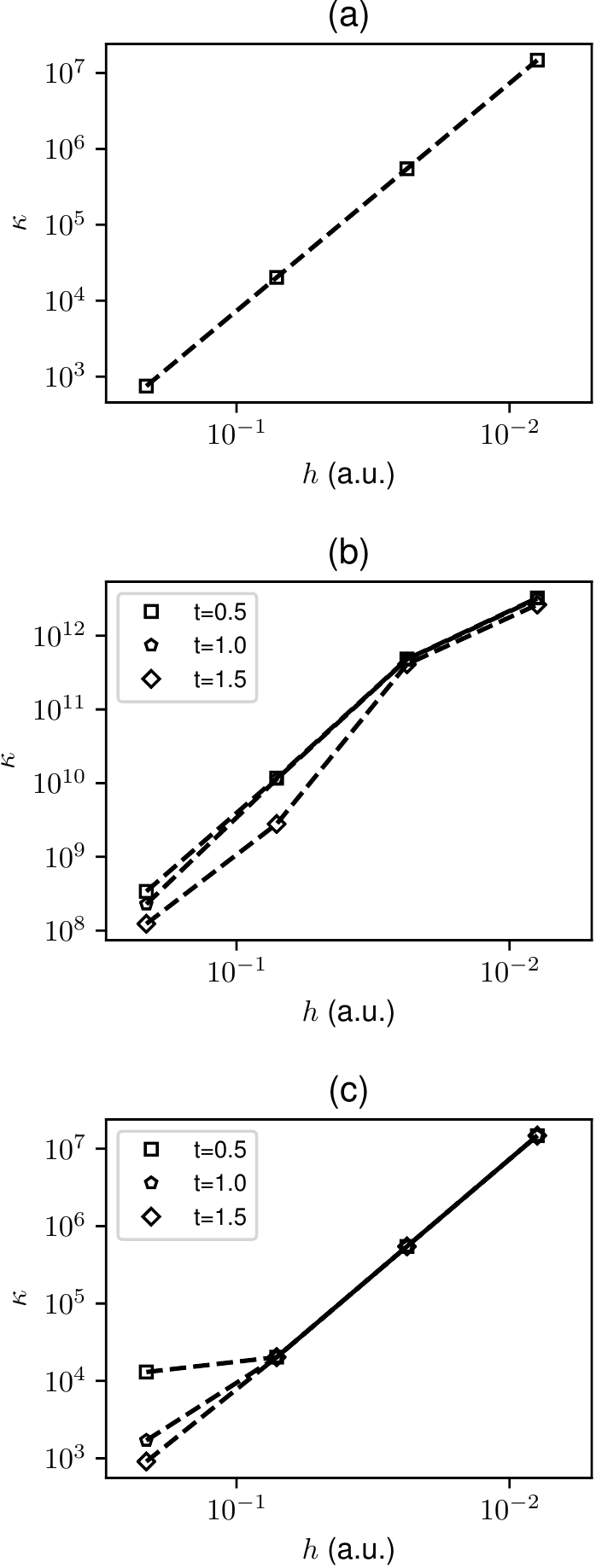}
    %\vspace{1cm}
    \caption{Condition number ($\kappa$) of the overlap matrix with respect to mesh size ($h$) for (a) CFE basis (b) EFE basis, and (c) OEFE basis.}
    \label{fig:cond}
\end{figure}

We demonstrate the effect of the finite element mesh size ($h$) on the conditioning of the CFE, the EFE, and the OEFE basis. The EFE basis is constructed as discussed in Ref.~\cite{kanungo2017large}. To elaborate, for the EFE basis, the enrichment functions are taken to be same as the atomic part ($\shp{A,u_{\bk}}{\nu}$) of $\shp{O,u_{\bk}}{\nu}$ (i.e., $\shp{O,u_{\bk}}{\nu}$ without the orthogonalizing component). Additionally, for the
EFE and the OEFE basis, we also report the effect of the smoothness factor $t$ (defined in Eq.~\ref{eqnCutoffProperties}) on the conditioning of the basis. We assess the conditioning of the basis through the condition number, $\kappa$, of its overlap matrix (defined as the ratio of the highest to lowest eigenvalue of the matrix). Given that the EFE and the OEFE basis has a k-point dependence, we consider the overlap matrix for the
$\Gamma$-point as being representative of the conditioning. 
%The materials system considered here is a single silicon atom in a periodic box of length 8.23845 a.u.
The materials system considered here is an 8-atom silicon unit cell of lattice constant 10.26 a.u..
For each of the three types of basis, we construct four different meshes by progressively refining the mesh near the nucleus. In the case of the EFE and OEFE basis, we set the cutoff distance $r_0$ (defined in Eq.~\ref{eqnCutoffProperties}) to 1.2 a.u., for all the enrichment functions, and vary the smoothness factor $t$. The results are shown in
Fig.~\ref{fig:cond}. As is expected, for all the three basis, the condition number increases monotonically with increasing refinement.  Evidently, the condition number of the EFE discretization (Fig.~\ref{fig:cond}(b)) is a factor $10^5-10^6$ higher than that of both CFE and OEFE basis, reaching beyond $10^{10}$ even for moderately refined meshes. Note that, for a given mesh size, increasing $t$ does lower the condition number for the EFE basis, but the
improvement is only marginal. In the case of the OEFE basis (Fig \ref{fig:cond}(c)), while the condition number increases with mesh refinement, it remains of the same order as that of the CFE basis. Further, based on our numerical studies, we observe convergence in the ground-state energies well before the condition number approaches $10^6$. 

\begin{figure}[htbp]
    \centering
    \includegraphics[scale=0.95]{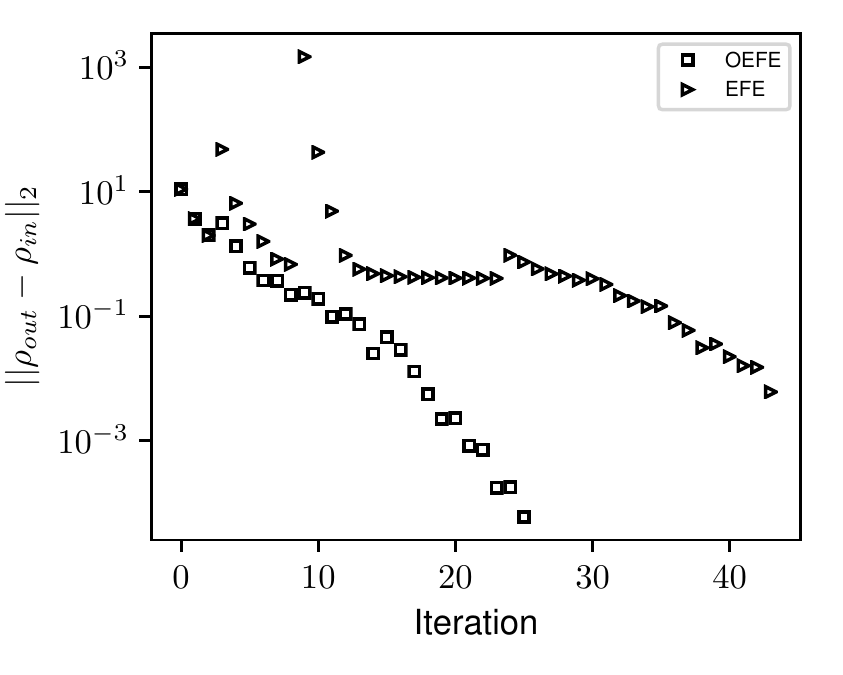}
    %\vspace{1cm}
    \caption{$L_2$ norm of the density difference with respect to SCF iteration number for the OEFE and the EFE basis.}
    \label{fig:illcondsic}
\end{figure}

In practical calculations, the implication of ill-conditioning may be one of following: (i) larger number of SCF iterations for convergence (ii) loss in accuracy, or (iii) failure to converge. To demonstrate this, we consider a $\Gamma-$point ground-state calculation on a 62 atom SiC divacancy system using both the EFE and the OEFE basis. We use the same underlying mesh for both EFE and OEFE calculations, and the resultant condition numbers of the overlap matrix in the EFE and OEFE cases are observed to be $10^{13}$ and $10^{7}$, respectively. Both  calculations use an n-stage Anderson mixing (mixing history of 20 and mixing parameter of 0.5) with a stopping criterion of $10^{-4}$ on the $L_2$ norm of the density difference. The variation of this norm with SCF iterations is plotted in Fig.~\ref{fig:illcondsic}. It is observed that self-consistency is reached in 25 iterations for the OEFE calculation while the EFE calculation struggles to converge even after 43 iterations. This demonstrates the importance of the OEFE basis for attaining accuracy and robustness in all-electron DFT calculations, while augmenting the CFE basis with enrichment functions.

\subsection{\label{sec:conv} Rate of convergence}
\begin{figure}[htp]
    \centering
    \includegraphics[scale=1.0]{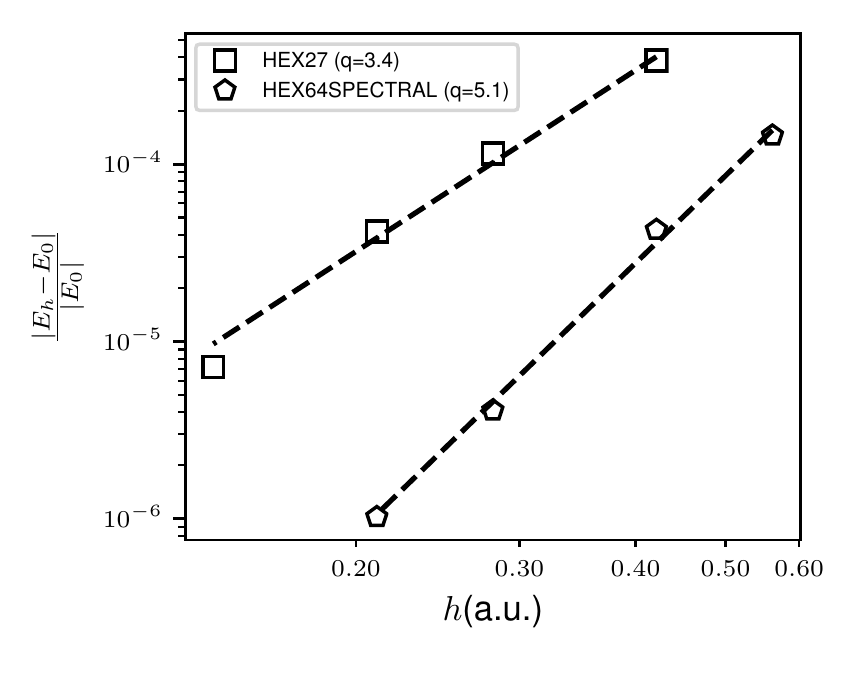}
    \caption{Convergence of ground-state energy with respect to element size for diamond.}
    \label{fig:c8}
\end{figure}

\begin{figure}[htp]
    \centering
    \includegraphics[scale=1.0]{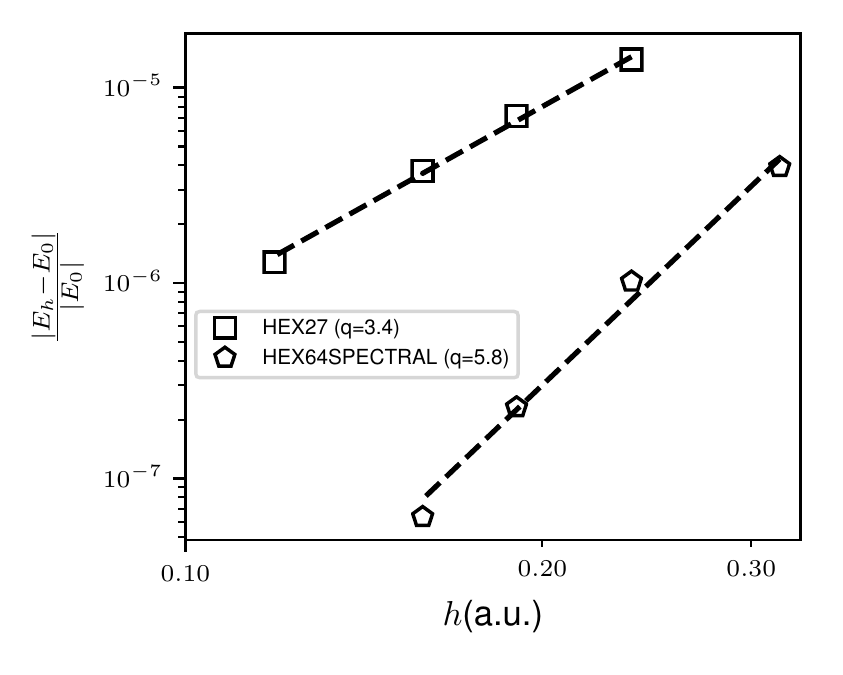}
    \caption{Convergence of ground-state energy with respect to element size for LiF.}
    \label{fig:lif}
\end{figure}

We now demonstrate the rate of convergence of the ground-state energy with respect to mesh refinement. The error in ground-state energy as a function of the mesh-size can be expressed as~\cite{Motamarri2013}  
\begin{equation}
    |E_h - E_0| = C h^{q}\,,
\end{equation}
where $E_h$ is the ground-state energy corresponding to a given finite element mesh of element size $h$, $E_0$ is the continuum ground-state energy corresponding to $h\rightarrow 0$, $C$ is a mesh-independent constant, and $q$ is the rate of convergence. We first evaluate $E_0$ using the OEFE basis with a highly refined higher-order CFE mesh. Subsequently, $C$ and $q$ are calculated by fitting the above relation to a given set of $E_h$ and $h$. As we are interested in studying the convergence with respect to discretization, we restrict these calculations to only $\Gamma$-point calculations.  We study the convergence on two materials systems: (i) an 8-atom carbon diamond-cubic system of lattice constant 6.74 a.u., and (ii) an 8-atom lithium fluoride (LiF) cubic halite system with lattice constant 7.6086 a.u.. For each system, we consider two types of finite elements---a quadratic finite element (HEX27) and a cubic spectral finite element (HEX64SPECTRAL). For each type of finite element, we construct a series of uniform meshes by refining the mesh-size $h$. Fig.~\ref{fig:c8} and Fig.~\ref{fig:lif} present the relative error in the energy as a function of the mesh-size for the diamond and the LiF systems, respectively. As evident, the numerical rates of convergence ($q$), reported in the figures, are in close agreement with the theoretical rate of $\mathcal{O}(h^{2p})$, where $p$ is the order of the finite element ($p=2$ for HEX27 and $p=3$ for HEX64SPECTRAL). The deviation from the theoretical rate is owing to errors that are beyond the basis discretization, i.e., errors due to quadrature, Chebyshev filtration tolerance, SCF convergence tolerance, etc. Furthermore, the $E_0$ per atom for the diamond system is -37.724793 Ha and is in close agreement with LAPW+lo value of -37.724827 Ha. Similarly, the $E_0$ per atom for the LiF system is -53.414248 Ha, which is again in good agreement with the LAPW+lo value of -53.414218 Ha.

\subsection{Unit cell calculations} \label{sec:unitcell}
\begin{figure}[htp]
    \centering
    \includegraphics[scale=0.9]{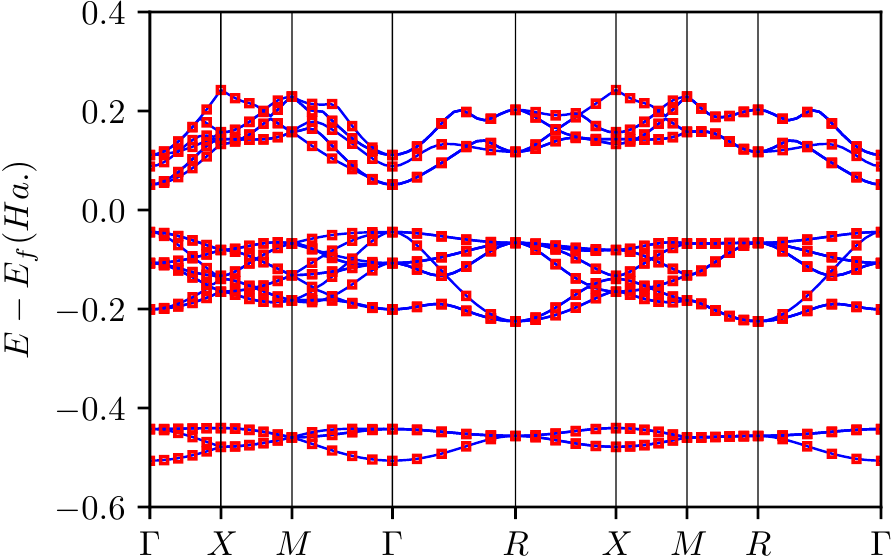}
    \caption{Bandstructure plot for MgS. Solid lines represent LAPW+lo data and points represent OEFE data.}
    \label{figbsMgS}
\end{figure}

\begin{figure}[htp]
    \centering
    \includegraphics[scale=0.9]{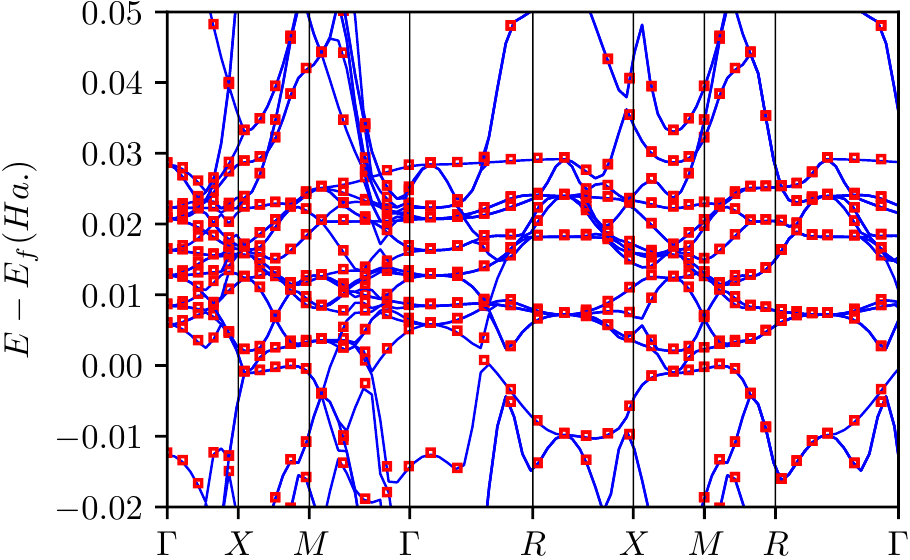}
    \caption{Bandstructure plot for Ce. Solid lines represent LAPW+lo data and points represent OEFE data.}
    \label{figbsCe}
\end{figure}

We assess the accuracy of the OEFE basis using two unit cell systems: (i) an 8-atom halite magnesium sulfide (MgS) with a lattice constant of 9.8266 a.u., and (ii) a 4-atom FCC cerium (Ce) unit cell with a lattice constant of 9.05 a.u.. We perform k-point converged ground-state calculations on both systems using the OEFE as well as the LAPW+lo basis. The ground-state energies for both MgS and Ce unit cells for different k-point grids are listed Table \ref{tableMgS} and Table \ref{tableCe}, respectively. As evident, the OEFE and LAPW+lo values agree to within 0.1~mHa. For MgS, we also show good agreement with the CFE based ground-state energy, evaluated at $\Gamma$-point. In case of Ce, a separate single atom in a box calculation (not shown in table) was performed to benchmark the accuracy of the OEFE basis with the CFE basis, and the results agree to within $1$~mHa. This was done since the 4-atom Ce calculations were prohibitively expensive with CFE basis. We also plot the bandstructure for the MgS (Fig.~\ref{figbsMgS}) and the Ce (Fig.~\ref{figbsCe}) near the Fermi level, obtained using OEFE and LAPW+lo basis. For both these materials systems, we see close agreement in the band structure obtained from the OEFE and the LAPW+lo calculations.

\begin{table}
\caption{Ground-state energy per atom (in Ha) of MgS unit cell for different k-point grid ($\Gamma$-point centered), using OEFE, LAPW+lo, and CFE basis.}
\begin{tabular}{c | c | c | c  }
k-pt & OEFE & LAPW+lo  & CFE   \\
\hline \hline
$1\times 1 \times 1$ & -298.06378 & -298.06383 & -298.06390  \\ 
$3\times 3 \times 3$ & -298.09548 & -298.09558 & n/a \\
$5\times 5 \times 5$ & -298.09556 & -298.09564 & n/a 
\end{tabular}
\label{tableMgS}
\end{table}

\begin{table}[h!]
\caption{Ground-state energy per atom (in Ha) of Ce unit cell for different k-point grid ($\Gamma$-point centered), using OEFE and LAPW+lo basis.}
\begin{tabular}{c|c|c}
k-pt & OEFE & LAPW+lo     \\
\hline \hline
$1\times 1 \times 1$ & -8563.72821 & -8563.72813  \\ 
$5\times 5 \times 5$ & -8563.63023 & -8563.63018 \\
$7\times 7 \times 7$ & -8563.63016 & -8563.63011 
\end{tabular}
\label{tableCe}
\end{table}

% \begin{table}
% \caption{Computational cost per SCF (in node-hr) for various SiC supercells, using OEFE (OEFE), LAPW+lo, and classical FE (CFE) basis.}
% \begin{tabular}{c | c | c | c }
% Supercell & OEFE & LAPW+lo & CFE    \\
% \hline \hline
% $2\times 2\times 2$ & 0.08  &  \textcolor{blue}{0.03} & 10.9 \\ 
% $3\times 3\times 3$ & 0.76 & \textcolor{blue}{0.73} & - \\
% $4\times 4\times 4$ & 6.4  & - & - \\
% $5\times5\times 5$ & 54 & - & -
% \end{tabular}
% \label{tableSiCPerf}
% \end{table}

% \begin{table}
% \caption{Total computational cost per SCF (in node-hr) and number of SCF iterations in parens for various SiC supercells, using OEFE (OEFE), LAPW+lo, and classical FE (CFE) basis.}
% \begin{tabular}{c | c | c | c }
% Supercell & OEFE & LAPW+lo & CFE    \\
% \hline \hline
% $2\times 2\times 2$ & 2.2 (21)  &  1.4 (56) & 197 (18) \\ 
% $3\times 3\times 3$ & 23.8 (27) & 46.9 (64) & - \\
% $4\times 4\times 4$ & 235 (34)  & - & - \\
% $5\times5\times 5$ & 2571 (46) & - & -
% \end{tabular}
% \label{tableSiCPerfTotal}
% \end{table}

%%%%%%%%%%%%%%%%%%%%%%%%% SUPERCELL CALCS rewritten %%%%%%%%%%%%%%%%%%%%%%%
\subsection{\label{sec:supercellcalcsnew} {Super cell calculations}}
We now demonstrate the accuracy and efficiency of the OEFE basis for large-scale all-electron DFT calculations. Moreover, wherever possible, we compare the performance of the OEFE basis against the CFE and the LAPW+lo basis. We remark that calculations on large periodic systems are often required to study the properties of defects in crystalline materials, so as to avoid the spurious defect-defect interactions arising from periodic boundary conditions.

Our large scale calculations consist of varying supercells for four different material systems: (i) divacancy in silicon carbide (SiC); (ii) NV-diamond; (iii) monovacancy in copper (Cu) ;  and (iv) divacancy in silver chloride (AgCl). 
The SiC supercell is constructed by translating the 8-atom diamond-structure cubic SiC unit cell of lattice constant 8.23845 a.u.. Subsequently, the divacancy in SiC is created by removing a second-nearest-neighbour pair of Si and C atoms from the supercell. 
The NV-diamond system is constructed from a diamond supercell by replacing a nearest-neighbor pair of C atoms by a nitrogen atom and a vacancy. The lattice constant of the 8-atom cubic diamond unit cell is taken to be 6.74 a.u.. 
The monovacancy in Cu is created by removing an atom from a supercell that has been constructed by translating the 4-atom FCC cubic unit cell of lattice constant 6.8 a.u..
Lastly, the AgCl divacancy system is constructed from the AgCl supercell by removing a pair of nearest neighbour Ag and Cl atoms. The lattice constant of the cubic hallite-structured AgCl unit cell is taken to be 10.3 a.u.. The supercell sizes used for each of these four systems are listed in Tables~\ref{tableSiCGSEnergy}--\ref{tableAgClGSEnergy}.

For each materials system, the basis is selected such that the discretization error in the ground-state energy for the $2\times2\times2$ supercell is less than $1$ mHa per atom. For the OEFE and the CFE basis, this amounts to using appropriately refined meshes and finite element orders. The approximate mesh sizes used near the nucleus and away from the nucleus and the finite element order used for the four materials systems have been listed in Table~\ref{tableMeshSize}. The discretization error in the LAPW+lo basis, on the other hand, is influenced by the \emph{rgkmax} (the product of the minimum muffin-tin radius and the maximum planewave cutoff), the linearization energies, muffin-tin radius, matching conditions at the muffin-tin boundary etc. In all but the Cu monovacancy calculations, the smallest possible rgkmax is selected to keep errors below $1$ mHa per atom. Default values were used for all other basis parameters as prescribed in the Elk code's species file. In the case of Cu monovacancy, however, default basis parameters lead to large errors and hence the \emph{highq} parameter set was used. The \emph{highq} parameter set in the Elk code, improves the accuracy of the calculation by boosting all basis-related parameters, including the rgkmax, from the default values. The LAPW+lo basis parameters used for the four material systems are summarized in Table~\ref{tableLapwParameters}. A major difference between the OEFE and the LAPW+lo basis based calculations lies in their treatment of the core electrons. The LAPW+lo employs a core-valence split, wherein the core states are solved using 1D radial solves and only the valence states are solved in 3D. However, as a first implementation of the OEFE basis, we treat all the states on the same footing and solve them in 3D (Eq.~\ref{eqnKSEigGenDiscrete}). 

\begin{table}[h!]
{\caption{Minimum element size $\text{h}_\text{min}$ (a.u.), maximum element size $\text{h}_\text{max}$ (a.u.), finite element order (p) , Chebyshev polynomial degree (m) and largest eigenvalue $\varepsilon_{\text{max}}$ (Ha.) for OEFE and CFE calculations}
\begin{tabular}{c|c|c|c|c|c}
Calculation & $\text{h}_\text{min}$ & $\text{h}_\text{max}$ & p  & m   &   $\varepsilon_{\text{max}}$   \\
\hline \hline
SiC-divac (OEFE) & 0.25 & 0.7  & 4 & 150   & $\mathcal{O}(10^{3})$\\ 
SiC-divac (CFE) & 0.02 & 0.7 & 5 & 2000  &$\mathcal{O}(10^{6})$\\ 
Cu-monovac (OEFE) & 0.2 & 0.6 &  3 & 150  & $\mathcal{O}(10^{3})$\\ 
NV-diamond (OEFE) & 0.6 & 0.6  &  4   & 50  &$\mathcal{O}(10^{2})$\\ 
AgCl-divac (OEFE) & 0.3 & 0.8 &  4  & 150  &$\mathcal{O}(10^{3})$
\end{tabular}
\label{tableMeshSize}
}
\end{table}

\begin{table}[h!]
{\caption{Basis set type and \emph{rgkmax} for LAPW+lo calculations}
\begin{tabular}{c|c|c}
Calculation & Basis set &  \emph{rgkmax}  \\
\hline \hline
SiC-divac  & Default & 7.5  \\ 
Cu-monovac  & \emph{highq} & 8.0 \\ 
NV-diamond  &Default & 7.0  \\ 
AgCl-divac  & Default & 8.0 
\end{tabular}
\label{tableLapwParameters}
}
\end{table}

We note that while the OEFE and the CFE implementation rely on an $L_2$ norm difference between densities at successive iterations as a convergence criterion for the SCF, the LAPW+lo implementation in Elk code uses a root mean square error (RMSE) in the Kohn-Sham potential as the convergence criterion. Thus, to use a consistent metric for comparing the OEFE basis against the LAPW+lo basis, we use a ground-state energy difference (between successive iterations) of $10^{-6}$ Ha per atom as a convergence criterion for the SCF for all calculations reported in this section. The number of SCF iterations is also influenced by the eigensolve tolerance and the type of mixing scheme used. In the case of the OEFE/CFE, instead of resorting to a tolerance for the eigensolve, we simply use a fixed polynomial degree for Chebyshev filter, as listed in Table~\ref{tableMeshSize}. This is equivalent to having a progressively tighter eigensolve tolerance with SCF iterations. In the case of the LAPW+lo calculations, between the iterative and the direct eigensolver available in Elk, we have found the direct eigensolver to be more efficient. Hence, we use the direct eigensolver for all the  LAPW+lo based calculations. We use an n-stage Anderson mixing scheme, with a history size of 20 and mixing parameter of $0.5$, for the OEFE/CFE calculations, while the Elk code uses the  Broyden mixing scheme. Further, the k-point sampling is restricted to the $\Gamma-$point, which is a reasonable approximation for large periodic domains.

All the calculations, except the SiC divacancy systems, are performed on the University of Michigan Great Lakes cluster’s 36-core nodes. The SiC divacancy systems are performed on NERSC's 68-core Cori-KNL nodes. The OEFE calculations for the four SiC divacancy systems are performed using 3, 9, 30 and 60 nodes, respectively. Similalrly, the OEFE calculations for the Cu monovacancy systems are performed using 2, 6, 12  and 15 nodes, respectively. The OEFE calculations for the NV-diamond systems are performed using 1, 5 and 10 nodes, respectively. Lastly, the OEFE calculations for the  AgCl divacancy systems are performed using 5 and 18 nodes respectively. For most of the OEFE calculations, each compute core is assigned to an MPI rank. The LAPW+lo calculations, on the other hand, are run on a single node with the number of OpenMP threads set to the number of cores in the node. We note that the Elk code does not offer distributed memory parallelism within a k-point. This lack of distributed memory parallelism, in turn, limits the system sizes that can be handled by the Elk code.

We, first, compare the accuracy of the OEFE, CFE, and LAPW+lo basis, in terms of the ground-state energies. Tables ~\ref{tableSiCGSEnergy}--\ref{tableAgClGSEnergy} compare the ground-state energies for all the four materials systems using the OEFE, CFE, and LAPW+lo basis. Given the high computational cost associated with the CFE basis, we limit the CFE calculations only to the $2\times2\times2$ supercell in SiC divacancy system. As is evident, the OEFE and CFE basis agree to within 0.3 mHa for the $2\times2\times2$ SiC divacancy system. Furthermore, for the systems where the LAPW+lo calculations are feasible, the OEFE and LAPW+lo basis agree to $0.5$ mHa, underlining the accuracy of the OEFE basis even for large-scale systems.

\begin{table}
{\caption{Ground-state energy per atom (in Ha) of various SiC supercells with a divacancy, using OEFE, LAPW+lo, and CFE basis. All reported energies are evaluated at $\Gamma$-point.}
\begin{tabular}{c | p {2cm}| c | c | c }
Supercell & \centering Atoms (Electrons) & OEFE & LAPW+lo & CFE    \\
\hline \hline
$2\times 2 \times2$ & \centering 62 (620) & -163.1053 & -163.1054 &  -163.1056  \\ 
$3\times 3\times 3$ & \centering 214 (2,140)  &-163.1119 & -163.1117 & -  \\
$4\times 4\times 4$ &  \centering 510 (5,100) & -163.1133 & - & -   \\
$5\times 5\times 5$ & \centering 998 (9,980)  & -163.1135 & - & - 
\end{tabular}
\label{tableSiCGSEnergy}
}
\end{table}

\begin{table}
{\caption{Ground-state energy per atom (in Ha) of various NV-diamond supercells, using OEFE and LAPW+lo basis. All reported energies are evaluated at $\Gamma$-point.}
\begin{tabular}{c | p {2cm}| c | c  }
Supercell & \centering Atoms (Electrons) & OEFE & LAPW+lo     \\
\hline \hline
$2\times 2 \times2$ & \centering 63 (379) & -38.0520 & -38.0522    \\ 
$3\times 3\times 3$ & \centering 215 (1,291)  & -37.8716  & -37.8720  \\
$4\times 4\times 4$ &  \centering 511 (3,067) & -37.8276 &  -
\end{tabular}
\label{tableNVDGSEnergy}
}
\end{table}

\begin{table}
{\caption{Ground-state energy per atom (in Ha) of various Cu supercells with a monovacancy, using OEFE and LAPW+lo basis. All reported energies are evaluated at $\Gamma$-point.}
\begin{tabular}{c | p {2cm}| c | c  }
Supercell & \centering Atoms (Electrons) & OEFE & LAPW+lo     \\
\hline \hline
$2\times 2 \times2$ & \centering 31 (899) & -1637.9256 & -1,637.9252   \\ 
$3\times 3\times 3$ & \centering 107 (3,103)  & -1637.9297 &  -1637.9294 \\
$4\times 4\times 3$ &  \centering 191 (5,539) & -1637.9355 &  -1637.9352 \\
$4\times 4\times 4$ &  \centering 255 (7,395) & -1637.9351 & -  
\end{tabular}
\label{tableCuGSEnergy}
}
\end{table}

\begin{table}
{\caption{Ground-state energy per atom (in Ha) of various AgCl supercells with a divacancy, using OEFE and LAPW+lo basis. All reported energies are evaluated at $\Gamma$-point.}
\begin{tabular}{c | p {2cm}| c | c  }
Supercell & \centering Atoms (Electrons) & OEFE & LAPW+lo     \\
\hline \hline
$2\times 2 \times2$ & \centering 62 (1,984) & -2,826.9589 & -2,826.9584    \\ 
$3\times 3\times 3$ & \centering 214 (6,848)  & -2,826.9597   & -2826.9592  
%$4\times 4\times 4$ &  \centering 510 (16,320) &   &  -
\end{tabular}
\label{tableAgClGSEnergy}
}
\end{table}

We, next, compare the relative performance of the OEFE basis against the LAPW+lo basis for all the four systems. Tables~\ref{tableSiCPerf}, \ref{tableNVDPerf}, \ref{tableCuPerf}, and \ref{tableAgClPerf} list the total computational cost for a ground-state calculation for the SiC, NV-diamond, Cu, and AgCl systems, respectively. Given that the OEFE and LAPW+lo implementations use different mixing scheme and eigensolve tolerances, which in turn effect the number of SCF iterations, we also provide the per SCF iteration computational cost as well as the number of SCF iterations. Moreover, for a comparison of the OEFE and CFE basis, we also provide the computational cost incurred by the CFE basis for the $2\times2\times2$ SiC divacancy system.

\begin{table}
{\caption{Comparison of OEFE, LAPW+lo, and CFE basis for the ground-state calculation on various SiC supercells with a divacancy: total computational cost ($C$ in node-hrs), computational cost per SCF iteration ($c$ in node-hrs) and number of SCF iterations ($N$). The total computational cost ($C$) includes the pre-SCF initialization costs.}
\begin{tabular}{|p{1.4cm} | c | c | c | c | c | c|}
\hline
\multirow{2}{1.4cm}{Supercell} & \multicolumn{2}{c|}{OEFE} & \multicolumn{2}{c|}{LAPW+lo} & \multicolumn{2}{c|}{CFE}    \\
\cline{2-7}
& $C$ & ($c$, $N$) & $C$ & ($c$, $N$) & $C$ & ($c$, $N$) \\
\hline \hline
$2\times 2\times 2$ & $1.48$ & (0.08, 12) & 1.28 & (0.04, 32) & 197 & (10.9,18) \\
\hline
$3\times 3\times 3$ & 13.92 & (0.76, 14) & 45.5 & (1.23, 37) & - & -\\
\hline
$4\times 4\times 4$ & 132.6  & (6.4, 18) & - & - & - & -\\
\hline
$5\times 5\times 5$ & 1102.5  & (45.9, 21) & - & - & - & -\\
\hline
\end{tabular}
\label{tableSiCPerf}
}
\end{table} 

\begin{table}
{\caption{Comparison of OEFE and LAPW+lo basis for the ground-state calculation on various NV-diamond supercells: total computational cost ($C$ in node-hrs), computational cost per SCF iteration ($c$ in node-hrs), and number of SCF iterations ($N$). The total computational cost ($C$) includes the pre-SCF initialization costs.}
\begin{tabular}{|p{1.4cm} | c | c | c | c |}
\hline
\multirow{2}{1.4cm}{Supercell} & \multicolumn{2}{c|}{OEFE} & \multicolumn{2}{c|}{LAPW+lo}     \\
\cline{2-5}
& $C$ & ($c$, $N$) & $C$ & ($c$, $N$)  \\
\hline \hline
$2\times 2\times 2$ & 0.19 & (0.008,12 ) & 0.32 & (0.02, 16)  \\
\hline
$3\times 3\times 3$ & 1.6  & (0.071,16 ) & 15.1 & (0.84, 18) \\
\hline
$4\times 4\times 4$ & 16.1  & (0.46,31 ) & - & - \\
\hline
\end{tabular}
\label{tableNVDPerf}
}
\end{table} 

\begin{table}
{\caption{Comparison of OEFE and LAPW+lo basis for the ground-state calculation on various Cu supercells with a monovacancy: total computational cost ($C$ in node-hrs), computational cost per SCF iteration ($c$ in node-hrs), and number of SCF iterations ($N$). The total computational cost ($C$) includes the pre-SCF initialization costs.}
\begin{tabular}{|p{1.4cm} | c | c | c | c |}
\hline
\multirow{2}{1.4cm}{Supercell} & \multicolumn{2}{c|}{OEFE} & \multicolumn{2}{c|}{LAPW+lo}     \\
\cline{2-5}
& $C$ & ($c$, $N$) & $C$ & ($c$, $N$)  \\
\hline \hline
$2\times 2\times 2$ & 0.92 & (0.033, 24) & 0.145 & (0.004, 32)  \\
\hline
$3\times 3\times 3$ & 20.6  & (0.55, 36) & 6.46 & (0.144, 45) \\
\hline
$4\times 4\times 3$ & 93.12  & (2.6, 35) & 50.63 & (0.92,55) \\
\hline
$4\times 4\times 4$ & 250.0  & (6.0, 41) & - & - \\
\hline
\end{tabular}
\label{tableCuPerf}
}
\end{table} 

\begin{table}
{\caption{Comparison of OEFE and LAPW+lo basis for the ground-state calculation on various AgCl supercells with a divacancy: total computational cost ($C$ in node-hrs), computational cost per SCF iteration ($c$ in node-hrs), and number of SCF iterations ($N$). The total computational cost ($C$) includes the pre-SCF initialization costs.}
\begin{tabular}{|p{1.4cm} | c | c | c | c |}
\hline
\multirow{2}{1.4cm}{Supercell} & \multicolumn{2}{c|}{OEFE} & \multicolumn{2}{c|}{LAPW+lo}     \\
\cline{2-5}
& $C$ & ($c$, $N$) & $C$ & ($c$, $N$)  \\
\hline \hline
$2\times 2\times 2$ & 2.17 & ( 0.25 ,7 ) & 0.24 & (0.015, 16)  \\
\hline
$3\times 3\times 3$ & 70.8  & ( 9.0, 7 ) & 9.8 & (0.58, 17) \\
\hline
%$4\times 4\times 4$ &   & (, ) & - & - \\
%\hline
\end{tabular}
\label{tableAgClPerf}
}
\end{table} 

\begin{figure}[H]
    \centering
    \includegraphics[scale=1.04]{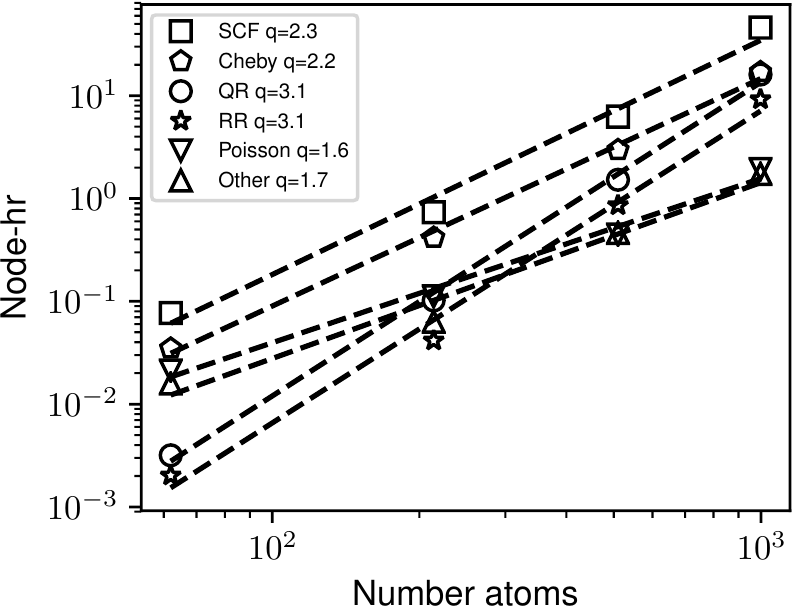}
    \caption{Computational cost (node-hr) per self-consistent field (SCF) iteration whose constituents include Cheby: Chebyshev filtration; QR: QR factorization; RR: Rayleigh-Ritz step (Projection + Direct diagonalization + Rotation); Poisson: Electrostatic Poisson problem; Other: Other costs including density calculation and Hamiltonian matrix construction.}
    \label{figcompcomplexSiC}
\end{figure}

The following observations can be made from the tables showing computational costs:
\begin{itemize}
    \item It is evident from the SiC divacancy $2\times2\times2$ calculations (cf. Table~\ref{tableSiCPerf}) that the OEFE basis is $130\times$ faster than the CFE basis. This staggering speedup is owing to a $\sim 15\times$ and a $\sim13\times$ reduction in the number of basis functions and Chebyshev polynomial degree, respectively (cf. Table~\ref{tableMeshSize}).
    
    \item For moderate system sizes, the OEFE basis outperforms the LAPW+lo basis for the SiC divacancy system and the NV-diamond system (cf. Tables~\ref{tableSiCPerf} and ~\ref{tableNVDPerf}) by a factor $3-9$.  
    
    \item For the Cu monovacancy system, the OEFE basis is $2\times$ slower than the LAPW+lo, for the largest comparable system (cf. Table~\ref{tableCuPerf}). In case of the AgCl divacancy system, the LAPW+lo basis significantly outperforms the OEFE basis (cf. Table~\ref{tableAgClPerf}).
    
    We remark that this comparatively inferior performance of the OEFE basis for systems with heavier atoms can be substantially improved by incorporating a core-valance splitting approach. To elaborate, while the LAPW+lo basis has the ability to split the spectrum into core and valence states and solve for the core states using 1D radial solves, in the OEFE basis, all states are treated on the same footing and are solved in 3D using Eq.~\ref{eqnKSEigGenDiscrete}. However, substantial speedup for OEFE basis, especially for systems with heavier atoms (where most of the states can be treated as core), can be realized by employing a spectrum splitting approach~\cite{Motamarri2017}, wherein a core-valence split can be attained by decomposing the occupied eigenspace into core and valence subspaces.
    
    % \tcr{This difference can be attributed to the ability of the  LAPW+lo basis to split the spectrum into core and valence states and solve for the core states using 1D radial solves.  On the other hand, in the OEFE basis, all states are treated on the same footing and are solved using Eq.~\ref{eqnKSEigGenDiscrete}. Hence, for heavier atoms with many localized (core) states, the LAPW+lo basis is expected to be more efficient as compared to the OEFE basis. For instance, in the case of Ag atoms, of the total 47 electrons, 30 electrons are treated as core states in the LAPW+lo basis.}
    
    % \tcb{While the OEFE remains slower than the LAPW+lo basis for the heavier systems, its performance can be improved by introducing a spectrum splitting approach~\cite{Motamarri2017}, wherein a core-valence split can be attained by decomposing the occupied eigenspace into a core and a valence subspace and projecting the Kohn-Sham Hamiltonian into the valence-subspace to evaluate the density matrix (and hence, the density). Given that for heavier atoms most of the electrons can be treated as core, a spectrum splitting approach can substantially boost the performance of the OEFE basis.}
    
    \item In terms of scaling with number of electrons ($N_e$) (i.e., weak scaling), the OEFE scales sub-cubically, in terms of the computational cost for an SCF iteration. To elaborate, we attain a scaling of $\mathcal{O}(N_e^{2.3})$, $\mathcal{O}(N_e^{2.0})$, $\mathcal{O}(N_e^{2.4})$ 
    for the SiC divacancy, NV-diamond, Cu monovacancy systems, respectively, even while accounting for system sizes ranging up to 9,980 electrons. This sub-cubic scaling is obtained because, in the regime of the system sizes considered, the dominant cost in OEFE calculation is the Chebyshev filtration step, which scales quadratically with the number of electrons. This is shown in greater detail for the SiC divacancy system in Fig.~\ref{figcompcomplexSiC}, where the scaling of various parts of the SCF algorithm are presented. In contrast, the scaling of LAPW+lo in Elk code is almost cubic even at smaller system sizes---$\mathcal{O}(N_e^{2.8})$, $\mathcal{O}(N_e^{3.0})$, $\mathcal{O}(N_e^{3.0})$
    for the SiC divacancy, NV-diamond, Cu monovacancy systems, respectively.
    
    \item Large system sizes are inaccessible using the implementation of the LAPW+lo basis in the Elk code, owing to memory limitations or impractical wall-clock times. The OEFE basis, on the other hand, is amenable to parallel implementation making large calculations possible within reasonable wall-clock times.
\end{itemize}

The above results, from the four material systems, underline the efficiency and robustness of the OEFE basis for large-scale all-electron DFT calculations, in comparison to both the CFE and the LAPW+lo basis.

%%%%%%%%%%%%%%%%%%%%%%%%% PARALLEL EFF %%%%%%%%%%%%%%%%%%%%%%%
\subsection{\label{sec:pareff} Parallel efficiency}
\begin{figure}[H]
    \centering
    \includegraphics[scale=1.1]{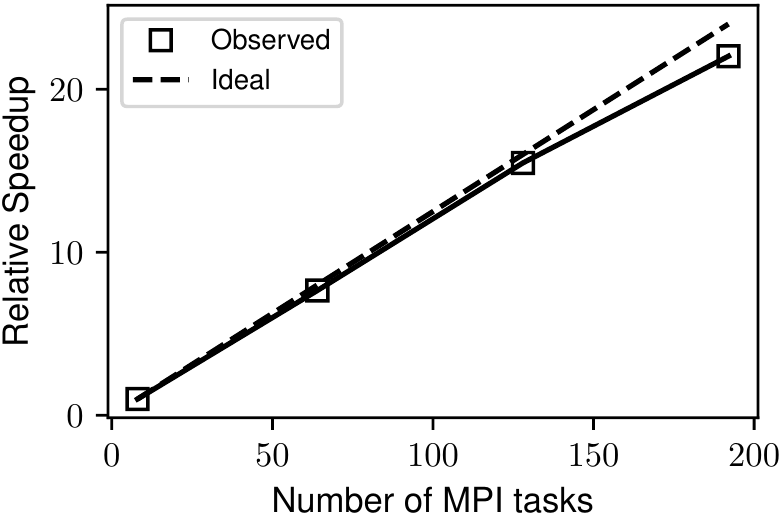}
    \caption{Strong scaling efficiency of the numerical implementation of OEFE basis using the $2\times 2\times 2$ SiC divacancy benchmark system.}
    \label{fig:pareff}
\end{figure}

We now present the strong scaling efficiency of our implementation of the OEFE basis using the $2\times 2\times 2$ SiC divacancy system. We used a discretization consisting of $\sim 1$ million CFE basis functions, and 434 and 62 orthogonalized enrichment functions for $\ualbk^h(\bx)$ and $\phi_{\text{aux}}^h$, respectively. The calculation is performed on increasing number of MPI tasks, ranging from 8 MPI tasks to 192 MPI tasks. The parallel efficiency is measured using the speedup relative to 8 MPI tasks, and is presented in Fig.~\ref{fig:pareff}. We observe $22 \times$ speedup with a parallel efficiency of 92\% at 192 MPI tasks, which demonstrates the good parallel scaling afforded by the formulation and the numerical implementation of our OEFE basis.

%%%%%%%%%%%%%%%%%%%%%%% SUMMARY %%%%%%%%%%%%%%%%%%%%%%%%%
%\input{summary}
\section{Summary} \label{sec:summary}
We have presented a systematically convergent and efficient basis, termed orthogonalized enriched finite element (OEFE) basis, for all-electron DFT calculations by augmenting the classical finite element (CFE) basis with enrichment functions constructed from single-atom Kohn-Sham orbitals and electrostatic potentials. In particular, we modify our previous formulation of the enriched finite element (EFE) basis~\cite{kanungo2017large} to alleviate the issue of potential ill-conditioning in the EFE basis. The key idea involved is to orthogonalize the enrichment functions with respect to the underlying CFE basis, while simultaneously maintaining the locality of the resultant basis. Additionally, we have optimized the basis for periodic calculation by introducing a k-point dependence to the enrichment functions. The resulting orthogonalized enrichment functions largely captures the sharp features of the electronic fields near the nuclei, reducing the requirement of a highly refined finite element mesh. This work establishes substantial computational advantage afforded by the OEFE basis over the CFE basis. 

In terms of robustness, the OEFE basis attained a significantly lower condition number of the overlap matrix compared to the EFE basis, while targeting the same chemical accuracy. The lower condition number, in turn, lends more efficiency to the OEFE basis by expediting the convergence of the SCF. Additionally, we demonstrated close to optimal rates of convergence for the ground-state energy with respect to the finite element mesh size, thereby underlining the systematic convergence (completeness) afforded by the OEFE basis. We established the accuracy of the  OEFE basis by attaining excellent agreement in ground-state energy and band structure with LAPW+lo method on benchmark calculations. Furthermore, we assessed the performance of the OEFE basis against the CFE and the LAPW+lo basis using increasing supercell sizes for four different materials system: (i) divacancy in SiC; (ii) NV-diamond, (iii) monovacancy in Cu; and (iv) divacancy in AgCl. For the system sizes accessible to the CFE basis, the OEFE basis attained a marked $130\times$ speedup. Further, the OEFE basis outperforms the LAPW+lo basis, for the moderate system sizes of the SiC divacancy and NV-diamond supercells considered in the study. However, the OEFE basis remains slower than the LAPW+lo basis for systems with heavier atoms---Cu monovacancy and AgCl divacancy supercells. We expect the performance of the OEFE basis for heavier atoms to improve substantially with the incorporation of an appropriate core-valence splitting~\cite{Motamarri2017} approach. Notably, using the OEFE basis we were able to conduct large-scale calculations on the SiC divacancy supercells, the NV-diamond supercells, and the Cu monovacancy supercells, with the largest system having 9,980 electrons. In contrast, LAPW+lo calculations on such large systems remained infeasible, owing to the parallel scaling and memory limitations of the LAPW+lo implementation in Elk code. Furthermore, within the benchmark systems considered, we attained a sub-cubic scaling with respect to the number of electrons, even accounting for system sizes ranging up to 9,980 electrons. In contrast, LAPW+lo basis in Elk exhibited cubic-scaling, even on small-to-moderate system sizes. Thus, the OEFE exhibits a later onset of the cubic scaling regime, as compared to the LAPW+lo basis. Lastly, we demonstrated close to ideal parallel scaling of our OEFE basis implementation up to $\sim200$ MPI tasks, for a 62 atom SiC divacancy system.

Thus, the proposed OEFE basis offers a robust, efficient, systematically convergent, and scalable basis for all-electron DFT calculations, applicable to metallic and non-metallic systems. Further improvement in the performance of the OEFE basis for systems with heavier atoms can be achieved by incorporating a core-valence spectrum splitting approach~\cite{Motamarri2017}. The use of the OEFE basis for all-electron time-dependent density functional theory (TDDFT) calculations~\cite{Marques2006, Kanungo2019} holds good promise, and is currently being investigated. Given, the importance of relativistic effects in all-electron calculations, an extension of this work to include both scalar relativistic and spin-orbit coupling effects constitutes a future direction of our research. Additionally, the OEFE ideas, in conjunction with the incorporation of configurational forces~\cite{Motamarri2018}, offers a powerful tool for all-electron Born-Oppenheimer molecular dynamics as well as Ehrenfest dynamics, and form an active line of our research. The OEFE basis can also offer a systematically convergent and efficient basis for the solution of the inverse DFT problem to compute the exact exchange-correlation potentials from ab-initio correlated densities~\cite{KZG2019}, and presents a worthwhile direction to pursue. Lastly, the proposed basis offers an efficient and accurate approach to treat the interaction between electronic and nuclear spins, which typically warrant all-electron calculations~\cite{Ghosh2019}.

%%%%%%%%%%%%%%%%%%%%% ACKNOWLEDGEMENT %%%%%%%%
%\input{acknowledgement}
\acknowledgements
{We gratefully acknowledge the support from the Department of Energy, Office of Basic Energy Sciences, grant number DE-SC0017380, under the auspices of which this work was conducted. This research used resources of the National Energy Research Scientific Computing Center, a DOE Office of Science User Facility supported by the Office of Science of the U.S. Department of Energy under Contract No. DE-AC02-05CH11231. V.G. also acknowledges the support of the Army Research Office through the DURIP grant W911NF1810242, which also provided the computational resources for this work.
}

\bibliography{ref}
\bibliographystyle{apsrev4-1}

\end{document}